\documentclass[usegraphicx,useAMS,usenatbib]{mn2e}
\usepackage{longtable,rotating,graphicx,graphics,subfigure,epsfig,epsf,amsmath,amssymb}

\newcommand{\degree}{$^{\circ}$}

\newcommand{\etal}{{\it et al.}}

\newcommand{\perbeam}{beam$^{-1}$}
\title[{\it HeViCS XII} ] {The Herschel Virgo Cluster Survey XII: FIR properties of optically-selected Virgo Cluster galaxies}
\author[Auld \etal]{R. Auld$^{1}$,
S. Bianchi$^{2}$,
M. W. L. Smith$^{1}$,
J. I. Davies$^{1}$,
G. J. Bendo$^{4}$,\newauthor
S. di Serego Alighieri$^{2}$,
L. Cortese$^{5}$,
M. Baes$^{3}$, 
D. J. Bomans$^{13}$,
M. Boquien$^{7}$, 
A. Boselli$^{7}$, \newauthor
L. Ciesla$^{7}$,
M. Clemens$^{9}$,
E. Corbelli$^{2}$,
I. De Looze$^{3}$,
J. Fritz$^{3}$, 
G. Gavazzi$^{8}$, \newauthor
C. Pappalardo$^{2}$,
M. Grossi$^{6}$,
L. K. Hunt$^{2}$,
S. Madden$^{10}$,
L. Magrini$^{2}$, 
M. Pohlen$^{1}$, \newauthor
J. Verstappen$^{3}$ 
C. Vlahakis$^{11,12}$.
E. M. Xilouris$^{14}$, 
S. Zibetti$^{2}$ \\
$^{1}$School of Physics and Astronomy, Cardiff University, The Parade, Cardiff, CF24 3AA, UK. \\
$^{2}$INAF-Osservatorio Astrofisico di Arcetri, Largo Enrico Fermi 5, 50125 Firenze, Italy. \\
$^{3}$Sterrenkundig Observatorium, Universiteit Gent, Krijgslaan 281 S9, B-9000 Gent, Belgium. \\
$^{4}$UK ALMA Regional Centre Node, Jodrell Bank Centre for Astrophysics, School of Physics and Astronomy, University of Manchester, \\
Oxford Road, Manchester M13 9PL, UK. \\
$^{5}$European Southern Observatory, Karl-Schwarzschild Str. 2, 85748 Garching bei Muenchen, Germany.  \\
$^{6}$ CAAUL, Observat\'orio Astron\'omico de Lisboa, Universidade de Lisboa, Tapada da Ajuda, 1349-018, Lisboa, Portugal. \\
$^{7}$Laboratoire d'Astrophysique de Marseille - LAM, Universit\'e d'Aix-Marseille \& CNRS, UMR7326,\\
38 rue F. Joliot-Curie, 13388 Marseille Cedex 13, France\\
$^{8}$Universita' di Milano-Bicocca, piazza della Scienza 3, 20100, Milano, Italy. \\ 
$^{9}$INAF-Osservatorio Astronomico di Padova, Vicolo dell'Osservatorio 5, 35122 Padova, Italy. \\
$^{10}$Laboratoire AIM, CEA/DSM- CNRS - Universit\'e Paris Diderot, Irfu/Service, Paris, France. \\
$^{11}$Joint ALMA Observatory, Alonso de Cordova 3107, Vitacura, Santiago, Chile\\
$^{12}$European Southern Observatory, Alonso de Cordova 3107, Vitacura, Casilla 19001, Santiago 19, Chile\\
$^{13}$Astronomical Institute, Ruhr-University Bochum, Universitaetsstr. 150, 44780 Bochum, Germany. \\
$^{14}$Institute of Astronomy and Astrophysics, National Observatory of Athens, I. 
Metaxaand Vas. Pavlou, P. Penteli, GR-15236 Athens, Greece. \\
$^{15}$Max-Planck-Institut fuer Astronomie, Koenigstuhl 17, D-69117 Heidelberg,  Germany. \\
 }

\begin{document}
\date{Submitted to MNRAS 2012 February 14}

\pagerange{\pageref{firstpage}--\pageref{lastpage}} \pubyear{2012}

\maketitle

\label{firstpage}

\begin{abstract}
The Herschel Virgo Cluster Survey (HeViCS) is the deepest, confusion-limited survey of the Virgo Cluster at far-infrared (FIR) wavelengths. The entire survey at full depth covers  $\sim$55 sq. deg. in 5 bands (100-500 \micron), encompassing the areas around the central dominant elliptical galaxies (M87, M86 \& M49) and extends as far as the NW cloud, the W cloud and the Southern extension. The survey extends beyond this region with lower sensitivity so that the total area covered is 84 sq. deg. In this paper we describe the data, the data acquisition techniques and present the detection rates of the optically selected Virgo Cluster Catalogue (VCC). We detect 254 (34\%) of 750 VCC galaxies found within the survey boundary in at least one band and 171 galaxies are detected in all five bands. For the remainder of the galaxies we have measured strict upper limits for their FIR emission. The population of detected galaxies contains early- as well as late-types although the latter dominate the detection statistics. We have modelled 168 galaxies, showing no evidence of a strong synchrotron component in their FIR spectra, using a single-temperature modified blackbody spectrum with a fixed emissivity index ($\beta = 2$). A study of the $\chi^2$ distribution indicates that this model is not appropriate in all cases, and this is supported by the FIR colours which indicate a spread in $\beta$=1--2. Statistical comparison of the dust mass and temperature distributions from 140 galaxies with $\chi^2_{dof=3} < 7.8$ (95\% confidence level) shows that late-types have typically colder, more massive dust reservoirs; the early-type dust masses have a mean of ${\rm log}(\langle M \rangle / M_{\sun}) = 6.3 \pm 0.3 $, while for late-types ${\rm log}(\langle M\rangle / M_{\sun}) =7.1 \pm 0.1$. The late-type dust temperatures have a mean of $\langle T\rangle=19.4\pm 0.2$ K, while for the early-types, $\langle T\rangle=21.1\pm 0.8$ K. Late-type galaxies in the cluster exhibit slightly lower dust masses than those in the field, but the cluster environment seems to have little effect on the bulk dust properties of early-types. In future papers we will focus more on the scientific analysis of the catalogue (e.g. measuring FIR luminosity functions, dust mass functions and resolved gas and dust properties).
\end{abstract}

\section{Introduction}

Of the major galaxy overdensities in the nearby Universe (Virgo, Fornax, Abell 1367, Ursa Major and Coma) the Virgo Cluster remains the most widely studied. Its proximity (17 Mpc, \citealt{gavazzi-1999}) allows us both to probe the cluster population down to faint levels and to resolve the larger galaxies with modest instruments.


In contrast to other, more relaxed clusters, Virgo exhibits a relatively high fraction of gas-bearing galaxies. As the galaxies surrounding Virgo fall into the cluster they are subject to many interactions. Interactions with the cluster gravitational potential (\citealt{merritt}, \citealt{boselli-2006}),  the hot intracluster medium (\citealt{gunn-gott}, \citealt{abadi}) and, in some cases, each other (\citealt{moore-1996}, \citealt{mihos-2004}) are among the dominant processes affecting evolution in the cluster. They are believed to be responsible for transforming galaxies from gas-bearing, star-forming late-types into the dormant early-type galaxies that dominate the cluster core, but the details of these processes are still poorly understood.

Evidence of these interactions can be seen in the disturbed morphologies of the galaxies themselves (e.g. NGC 4438, \citealt{oosterloo}, \citealt{vollmer}) or in the intracluster medium (ICM), where streams such as those observed in the optical \citep{mihos-2005} and 21cm \citep{gavazzi-alfa-hi} show that the space between galaxies is littered with stellar and gaseous material. The other major ISM constituent, dust, is also expected to carry the signatures of these interactions. Understanding what happens to the dust component in the cluster environment is another important aspect of environmental effects on galaxy evolution, but requires a large dataset of galaxies in different environments.


Dust emission dominates the mid infrared (MIR, 1-10 \micron) to far
infrared and sub-mm (FIR, 10-500 \micron; sub-mm, 500-1000 \micron)
portion of a galaxy's spectral energy distribution (SED). In the MIR,
it is a complicated mixture of thermal emission from stars and warm
(T$\sim$30 K) dust, and line emission from molecular species. The FIR
and sub-mm regime, however, is dominated by thermal emission from
interstellar dust grains. Studying galaxies at these wavelengths has
been problematic since ground-based observations are compromised by high attenuation and bright spectral line emission from the Earth's atmosphere. 

Space missions such as {\it IRAS}, {\it ISO} and most recently {\it Spitzer}, have been vital for expanding our understanding of interstellar dust. IRAS and Spitzer were sensitive to the warm dust ($>30$K) within dust-bearing galaxies (e.g. \citet{devyoung-iras-1990}, \citealt{sings}), while ISO and the ground-based SCUBA allowed astronomers to probe the coldest dust within galaxies (e.g. \citealt{popescu-iso}, \citealt{tuffs2002}, \citealt{alton-1998a}, \citealt{duneales-slugs-2001}, \citealt{duneales-slugs-2002}). Unfortunately these pioneering instruments were limited in their resolution and sensitivity, and these early studies were confined to individually targeted galaxies.

The {\bf He}rschel {\bf Vi}rgo {\bf C}luster {\bf S}urvey (HeViCS, \citealt{hevics1}) has observed 84 sq. deg. of the Virgo cluster with the primary goal of studying cold-dust bearing galaxies in the cluster environment. The 3.5m primary mirror of the Herschel Space Observatory, combined with the ability to perform 5-band photometry simultaneously over a wavelength range of 100--500 \micron, afford the survey unprecedented sensitivity and resolution in the FIR regime. This has enabled us to conduct the most detailed studies to-date of cold ($<30$ K) dust in Virgo cluster galaxies.

In previous work, based on shallow data from the central 4\degree\ $\times$ 4\degree\ region, we have already analysed the FIR luminosity function in the cluster core \citep{hevics1}, studied truncated dust disks in cluster galaxies \citep{hevics2}, put upper limits on the dust lifetime in early-type galaxies \citep{hevics3}, probed the distribution of dust mass and temperature within cluster spirals \citep{hevics4}, identified an excess at 500 \micron\ for a sample of metal-poor, star-forming dwarfs \citep{hevics5}, studied non-thermal FIR emission from the dominant cluster elliptical galaxy, M87 \citep{hevics6},  detected dust in cluster dwarf elliptical galaxies \citep{hevics7}.

In more recent work, based on the full survey area but limited in depth, we have presented an analysis of the brightest galaxies at 500\micron\  (BGS, \citealt{hevics8}). We have examined the metallicity dependence of the molecular gas conversion factor, $X_{CO}$ \citep{hevics9}, and probed the effect of interactions on the distribution of dust and gas in a sample of late-types \citep{hevics10, hevics11}, 

In this, the first paper to exploit the full dataset, we present the FIR properties of a statistically complete, optically-selected sample; the Virgo Cluster Catalogue (VCC, \citealt{bing-vcc}). We concentrate on the description of the data, the data reduction, the flux measurements themselves and the recovery rates of the VCC galaxies. This will pave the way for future papers which will focus on the science that can be extracted from the data (FIR/sub-mm colour relations, luminosity functions, dust mass functions, radial/resolved FIR properties etc.).

The VCC is the largest and most reliable catalogue of cluster members; it  covers 140 sq. deg. and contains almost 1300 probable cluster galaxies. Membership was based, for most part, on morphological criteria, with redshift information taking precedence as it became available. A comparison of the morphology method with redshift data proved it to be very reliable, with a success rate of over 98\% (\citealt{bing-vcc}, \citealt{bing-spec}). 

While it is true that the cluster has been observed with more sophisticated instruments with superior imaging capabilities compared to the original Las Campanas plates (e.g. \citealt{kim-sdss}, \citealt{cote}), the power of the catalogue lies in the membership information itself. Applying their optical completeness limit of photographic magnitude, $m_{pg} \lid 18$, yields a population of $\sim$1000 galaxies, and has provided the astronomical community with a large, statistically complete sample of Virgo Cluster galaxies. It has been used for nearly 30 years with only minor revision (e.g. \citealt{bing-spec}, \citealt{gavazzi-1999}) and will only be surpassed by the completion of the Next Generation Virgo Survey (\citealt{mei}, \citealt{ferrarese}).

The data for our optical comparisons were taken from the {\sc goldmine} database \citep{goldmine} which includes measurements at multiple wavelengths for the entire, revised VCC. It is our intention to expand the database with our FIR measurements. By augmenting these data with the HeViCS data we aim to provide the community with a definitive legacy product that will enable the detailed study of stars and dust in the cluster environment.

In section \ref{obsdr} we describe the survey data acquisition, data reduction, the flux extraction method and the SED fitting. In section \ref{results} we present the detection rates and the flux measurements at 100, 160, 250, 350 \& 500 \micron. In this section we also provide the results of the SED modelling and show the derived dust mass and temperature distributions for a subset of galaxies which were well fit by the single-temperature model. In section \ref{summary} we summarise our findings. The tables of fluxes and dust mass/temperature are attached as appendices in Tables \ref{fluxtbl} \& \ref{temptbl}. The FIR SED fits are displayed in Appendix \ref{onetemps}.

\section{Observations, Data Reduction and Flux Measurement}
\label{obsdr}
\subsection{Observations}
\begin{figure*}
\centering
	\includegraphics[height=0.89\textheight]{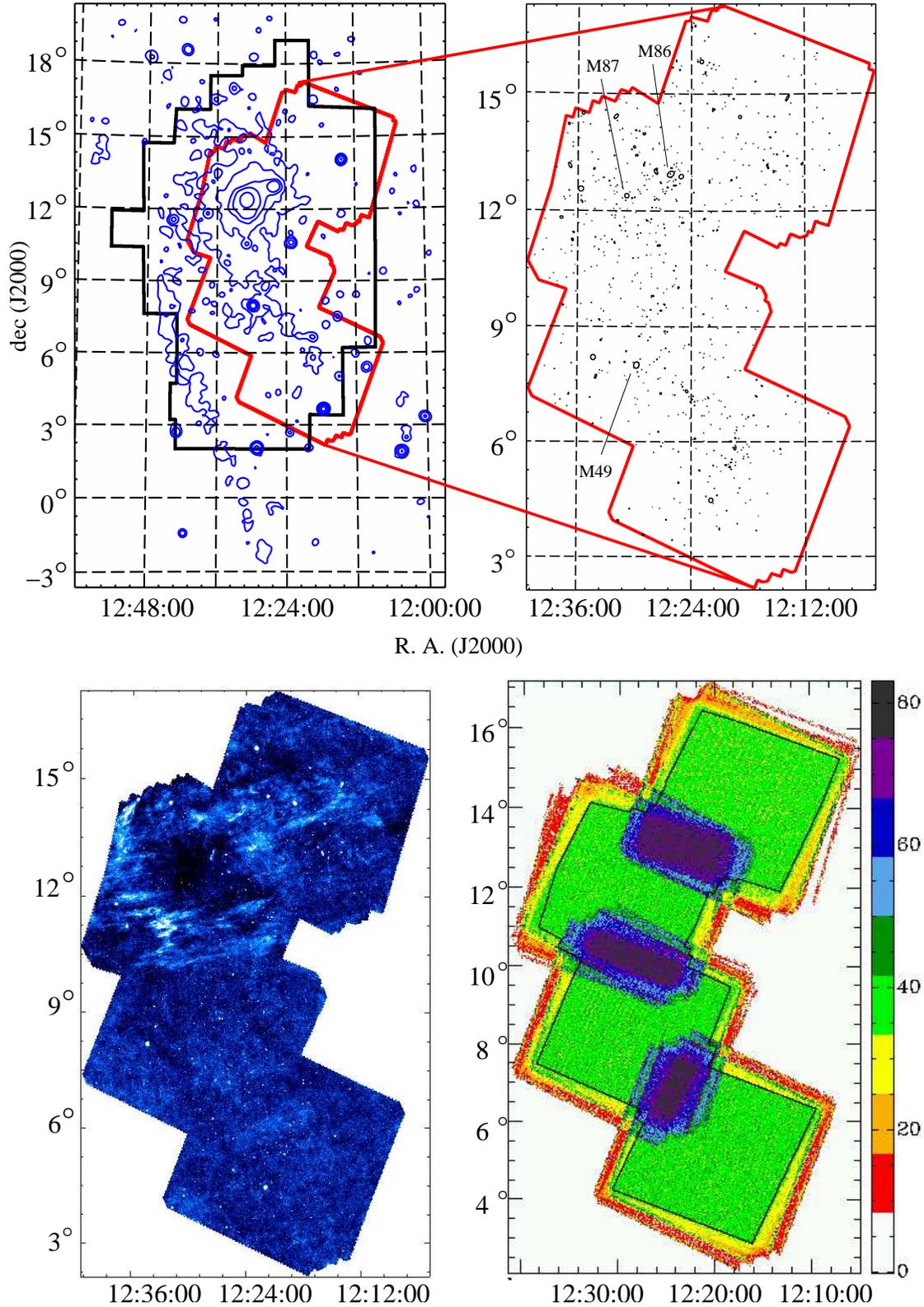}
	\vspace{0.0cm}
	\caption{{\it Top left}: the Virgo Cluster region. X-ray contours from Boehringer et al. are shown in blue. The VCC survey region is outlined in solid black and the full extent of HeViCS is outlined in red. {\it Top right}: the HeViCS survey region with black ellipses representing the VCC optical discs measured to $d_{25}$. The dominant cluster galaxies have been labelled and their positions correspond to the peaks in X-ray emission. {\it Bottom left}: {\it Herschel-SPIRE} 250\micron\ image of the full survey. Even in this small image it is possible to identify the VCC galaxies with the strongest FIR emission and the large swathes of Galactic dust cirrus. {\it Bottom right}: Survey depth, measured in samples per pixel, from the 250 \micron\ data. The inner region, covered by 8 scans with {\it PACS} and {\it SPIRE}, is shown by the black outline. In the overlap region between tiles the coverage rises to 16 scans.}
	\label{fig1}
\end{figure*}

The HeViCS survey region constitutes four overlapping 4\degr$\times$4\degr\  tiles (Fig. \ref{fig1}) including the majority of Cloud A, centred on M87 and extending far enough to encompass the NW cloud, the W cloud and the S extension in the nomenclature of \citet{devau1961}. Each tile has been observed by the Herschel PACS \citep{poglitsch} and SPIRE \citep{griffin-spire} instruments operating in fast-parallel mode, allowing PACS and SPIRE photometry to be performed simultaneously at a scan rate of 60\arcsec s$^{-1}$. The observing strategy consists of scanning each 4\degr$\times$4\degr\  tile in two orthogonal directions. This cross-linking allows for the application of data processing techniques that can mitigate the effects of $1/f$ noise. This enables the potential recovery of large scale structure (cirrus/intracluster dust) of up to the length of a tile (4\degr). This strategy is then repeated until each tile is nominally covered by eight scans, although in overlap regions this can reach as high as sixteen scans. 

The {\it PACS} footprint is offset from the SPIRE footprint by 20\arcmin, leading to slight mismatch in the area of sky surveyed. To compensate for this effect, the proposed survey target regions are automatically expanded at the observing stage, to reach full coverage with both instruments in the target field. This gives rise to extra data, outside the target region. We retain the extra data and after combining the entire set of cross-linked scans, the total sky coverage is somewhat larger than previously claimed ($\sim$84 sq. deg. compared to 64 sq. deg.), although the full-depth region is limited to the central 55 sq. deg.

\subsection{Data Reduction}
\label{dr}
The SPIRE photometer \citep{griffin-2010} data were processed up to Level 1 (i.e. calibrated bolometer timelines) with a custom made pipeline adapted from the official pipeline \footnote{See \citet{griffin-2010} or \citet{dowell-2010} for a more detailed description of the official pipeline and a list of the individual modules.}. The purpose of the pipeline is to remove all instrumental artefacts such as glitches, finite bolometer time response, electronic filtering and thermal drift, as well as applying astrometry and flux calibration.

The main difference between our pipeline and the standard one is that we did not run the default {\it temperatureDriftCorrection} and the residual, median baseline subtraction. Instead we use a custom method called the {\it BRIght Galaxy ADaptive Element } ({\bf BriGAdE}, Smith et al. 2012, in prep.) to remove the temperature drift. No further baseline subtraction was necessary to bring the bolometer baselines to a common level.

\begin{figure}
\centering
	\includegraphics[width=0.45\textwidth]{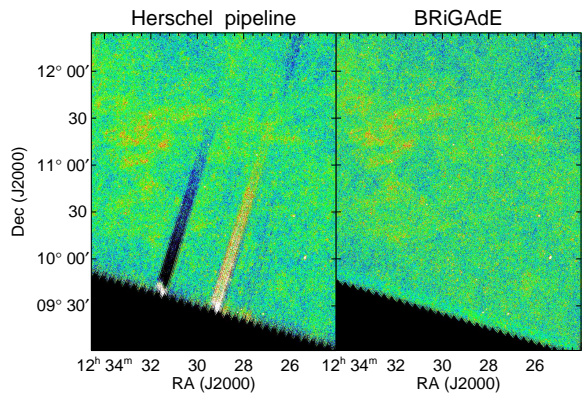}
	\caption{Instantaneous jumps in the thermistor timelines, give rise to striping in the final images, when using the standard Herschel pipeline. Use of the BriGAdE method greatly reduces their impact.}
	\label{fig1b}
\end{figure}

BriGAdE uses the information from the thermistors in each array and directly fits the thermistor data to the entire bolometer timeline (including data where the spacecraft was slewing between scans). If both thermistors exhibited instantaneous `jumps' (an artefact where there is a sudden offset in the timelines), these are either corrected or the comparison switched to the slightly less sensitive Dark Pixels of the individual array. This approach is hampered slightly by the presence of bright sources in the bolometer timelines. To suppress their influence, they are automatically removed from the fitting process along with samples affected by other artefacts (`jumps' and glitches). When a choice of thermistors is available (i.e. for the 250 \& 500 \micron\ arrays), the one providing the best fit is used to subtract a scaled version from the bolometer timelines. This method improves the baseline subtraction significantly, especially in cases where there are large or rapid temperature variations during the observations, which cause large stripes in the final maps (Fig. \ref{fig1b}).

The individual scans from all 32 cross-linked observations are combined into a single mosaic using the inbuilt na\"ive mapper of HIPE. The full-width half-maximum (FWHM) measurements of the beams in the final maps are 18.2\arcsec, 25.4\arcsec\  \& 36.0\arcsec\ for 250, 350 \& 500 \micron\ respectively. The beam areas used in the conversion from Jy \perbeam\ to Jy were 423, 751 and 1587 arcsec$^2$, at 250, 350 \& 500 \micron\ respectively. This is in accordance with the SPIRE photometry guidelines\footnote{http://herschel.esac.esa.int/twiki/pub/Public/ SpireCalibrationWeb/beam\_release\_note\_v1-1.pdf}. The final maps have pixel sizes of 6\arcsec,8\arcsec\ \& 12\arcsec\ at 250, 350 \& 500 \micron\ respectively. One of the by-products of the HIPE data processing is a map of the instrumental noise. For the SPIRE bands this gives corresponding instrument-associated noise levels of 4.9, 4.9 \& 5.7 mJy \perbeam\ in the areas uniformly covered by 8 scans. This drops to 3.5, 3.4, \& 4.0 mJy \perbeam, in the overlapping regions, uniformly covered by 16 scans. The values are consistent with a drop in the noise level proportional to $\sqrt(t)$ where $t$ is the integration time, demonstrating that the contamination by $1/f$ noise is minimal.

Deriving the global noise is not a straightforward task due to the presence of background source crowding and Galactic cirrus emission, which are both position-dependent. We selected a region in the fully-sampled southernmost tile, which had the lowest cirrus contribution. Within this region we measured pixel-pixel fluctuations and applied iterative 3-$\sigma$ clipping to eliminate the contribution from bright sources. This gave noise values of 6.6, 7.3 \& 8.0 mJy/beam (250, 350, \& 500 \micron\  respectively). By quadratically subtracting the instrumental noise, we derived confusion noise estimates of 4.4, 5.4 \& 5.6  mJy/beam (250, 350, \& 500 \micron\  respectively). These crude estimates are slightly higher than \citet{nguyen} on the Hermes fields (3.8, 4.6 \& 5.2 mJy/beam respectively, employing an analogous iterative 3-$\sigma$ clip), but still within twice their quoted uncertainty. The HeVICS SPIRE maps have similar contributions from confusion and instrumental noise. We repeated the analysis on regions of overlap between the tiles, which have greater coverage and found no significant reduction in the noise level, implying that the maps are close to being dominated by confusion noise.

With regard to the PACS data, we have obtained maps at 100 and 160 \micron\ by exploiting the naive projection task \texttt{photProject} in HIPE (v7.3.0), following a similar approach as described by \citet{ibar}. Map making was performed in two steps: for each single scan, we first obtained maps using the standard data reduction pipeline, but with the entire VCC galaxies masked out. In these maps, the $1/f$ noise was corrected for by applying a high-pass filter with a length of 20 and 40 frames for the 100 and 160 \micron\ map, respectively, and deglitching was performed by means of the sigma-clipping standard algorithm (\texttt{IIndLevelDeglitchTask} task in HIPE, with $\sigma=3$). These preliminary maps were used to identify and mask bright, extended non-VCC sources, and to fine-tune the deglitching. The masks were then applied to the calibrated timelines using a flux threshold set to $2.5\sigma$, and the minimum number of pixels above the threshold, for a source to be detected, was set to 48 and 24 for the 100 and 160 \micron\ channels, respectively. In order to minimize the high-pass filtering over-subtraction, the masks were enlarged, by adding 4 and 2 pixels to the edges of the detected sources, for 100 and 160 \micron\  maps, respectively. 

In a second step, the data were reduced again. This time, the deglitching task was applied before high-pass filtering, with the bright sources being masked to avoid unwanted flux removal from their brightest parts. Then, the masked, deglitched timelines were high-pass filtered, with a filter length of 10 and 20 frames for 100 and 160 \micron\, respectively. At this point the timelines for the eight scans were joined together. After this, the second-level deglitching task was run to remove glitches from the bright sources and, finally, individual tile maps were produced with the \texttt{photProject} task.

Due to the enormous volume of data in a PACS observation, the HIPE map maker is currently only capable of combining 8 cross-scans from an individual tile into a single mosaic. The individual tiles were then combined into mosaics using the {\it SWarp} \footnote{http://www.astromatic.net/software/swarp} package. 

The PACS beams at 100 and 160 \micron\  are elongated along the scan direction due to onboard averaging of the PACS data. This is a necessary complication due to the limit of the rate at which data can be downloaded from the spacecraft. 
Averaging together the 8-scan data, taken at different observing angles, results in approximately circular beams. These are Gaussian in shape with FWHM values for the 100 \& 160\micron\ beams of $9.4\arcsec$\  and $13.4\arcsec$\  respectively and the corresponding map pixel sizes are 2\arcsec\  and 3\arcsec. The PACS instrumental error maps from the HIPE pipeline are still in the process of being optimised, so PACS errors were estimated directly from the signal maps, using apertures on blank regions of sky. The standard deviations of the background in the PACS maps are 1.9 and 1.2 mJy pixel$^{-1}$ for the 160 and 100 \micron\ channels respectively in the areas covered by 8 scans. This decreases to 1.3 and 0.8 mJy pixel$^{-1}$ in the overlapping regions. This is consistent with a $\sqrt t_{int}$ decrease, where $t_{int}$ is the integration time, indicating that the noise in the background in the PACS maps is still dominated by instrumental noise.

\subsection{Flux Measurement}
\label{fluxmeas}
Of the 1076 VCC objects with $m_{pg} \lid 18$, 750 fall within the extreme boundary of the HeViCS region. Once these targets were identified, the optical parameters (position, ellipticity, position angle, and optical diameter -- $d_{25}$) were used to overplot the optical disk on the 250\micron\ image. Position angles for the VCC galaxies were obtained from HyperLEDA. The $d_{25}$ was taken from the original VCC as quoted in {\sc goldmine}, which has been estimated from the photographic plates. Previous studies (e.g. \citealt{pohlen}, \citealt{hevics2}) have shown that the optical emission is fairly well traced by the FIR emission, for late-type galaxies without truncated gas-disks. Preliminary comparisons between the optical and FIR emission for HeViCS late-type galaxies revealed that the optical ellipse parameters provided a good basis for defining the FIR apertures for measuring the flux. Early-types tended to exhibit point-like emission in the FIR, and in a few particular cases irregular and extended (e.g. M86) or dominated by synchrotron radiation from the AGN jets (e.g. M87 \citealt{hevics6}).

\begin{figure*}
	\includegraphics[angle=0,height=0.9\textheight]{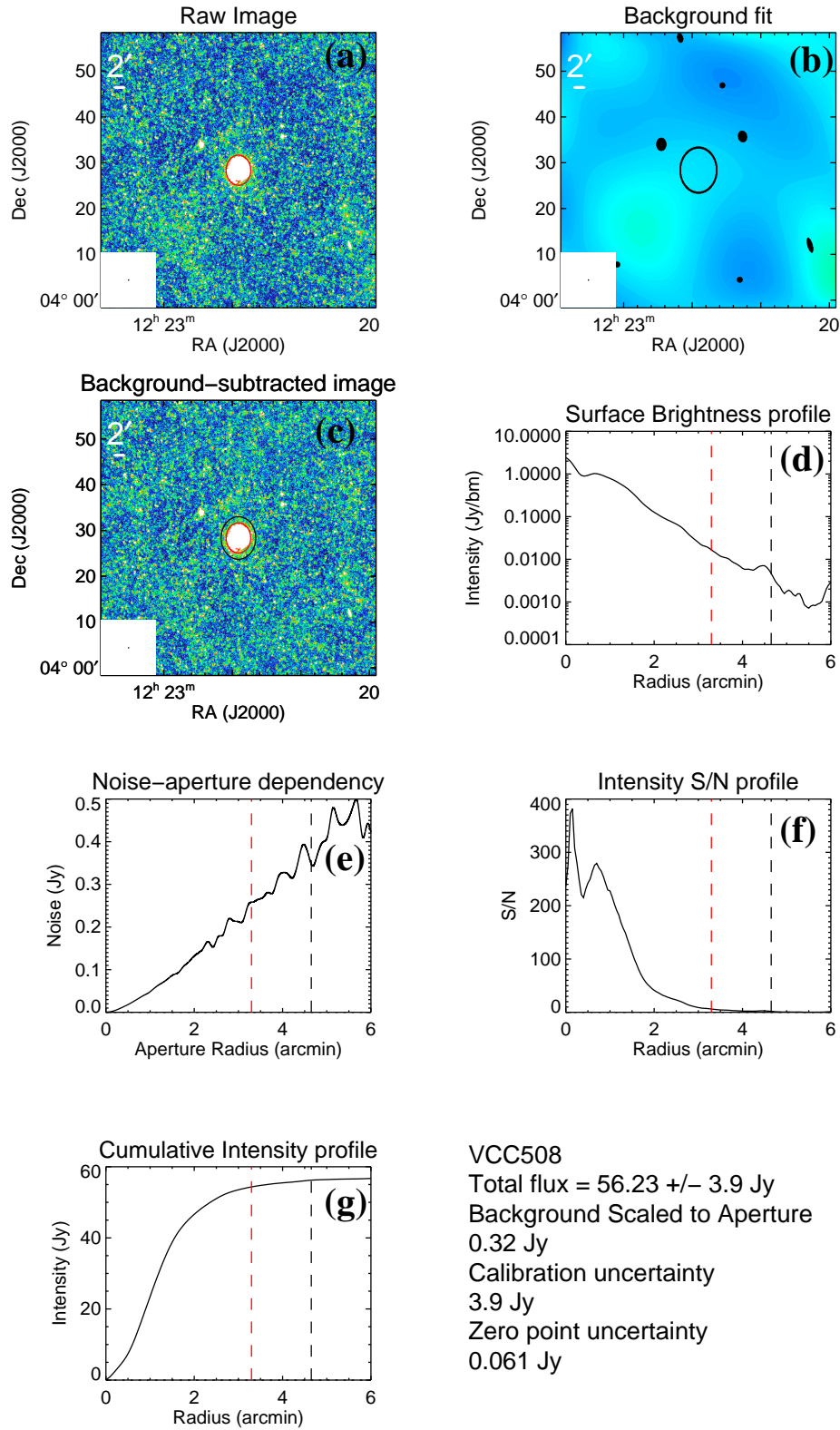}
	\caption{Example products from the fitting process for M61 (VCC508) at 250\micron. {\bf (a)} The raw sub-image extracted from the full HeViCS mosaic. The $d_{25}$ limit is outlined in red, the 250 \micron\ beam is given to scale at the bottom left of the image. {\bf (b)} The estimate of the background from the masked image. The ellipse is explained in the text. {\bf (c)} The resultant background-subtracted sub-image on which radial plots and flux measurements are performed. {\bf(d)} Radial plot of surface brightness. (e) radial dependence of the total noise in a circular aperture. {\bf (f), (g)} Radial plots of  S/N \& cumulative intensity respectively. Optical ($d_{25}$) and FIR limits are indicated by the red and black dashed lines respectively.\label{fig2}}
\end{figure*}

Accurate photometry had to be able to cope with several aspects inherent to the Herschel images. The SPIRE maps are dominated by Galactic cirrus and, to a lesser extent, by unresolved background sources. Some bright sources in the PACS images also suffer from negative bowls which are the result of inaccurate masking of bright sources during the high-pass filtering. Various stages of the measurement process are displayed in Fig. \ref{fig2}. It may aid the reader to refer to this during the description that follows.

To begin with, a 200$\times$200 pixel sub-image centred on each galaxy, was extracted from the full map (Fig. \ref{fig2}a). The galaxy was then masked as are all the nearby Virgo galaxies. The extent of the galaxy was defined by the $1.5\times d_{25}$ limit. It should be noted that there can be small differences between this $d_{25}$ measurement (estimated from photographic plates) and the equivalent measured from CCD data. The difference is not important to us, however, since the ellipticity and extent are merely used as starting points for the ellipse measuring procedure. To account for edge-on galaxies and those whose optical dimensions were smaller than the beam at any given wavelength, the optical ellipse parameters were convolved with a Gaussian beam profile of the relevant size for each band. In some cases, the optical morphology was unrepresentative of the FIR emission (e.g. for point-like or irregular sources). For bright, point-like sources, circular ellipses were employed and for M86, we used the features identified in \citet{gomez} to define the extent of emission in M86. For cases where the optical extent was too large to elicit a reliable estimate of the background, the sub-image region was increased to 600 pixels for SPIRE and up to 1200 pixels for PACS maps. 

The background is a combination of the foreground cirrus, bright background sources and unresolved background sources. For SPIRE images which are highly confused, the estimate of the background was achieved with a 5th order (i.e. terms up to $x^5, y^5$ plus all cross terms) 2-D polynomial over the entire masked sub-image. We employed the least-squares polynomial fitting {\sc idl} routine {\sc sfit} to perform the fit. To counteract the influence of bright background galaxies a 95\% flux clip was imposed and the remaining pixels used to estimate the background. For PACS images which were dominated by instrumental noise, a 2nd order polynomial was adequate. An example of the background estimate is shown in Fig. \ref{fig2}(b). Once the background estimation was obtained, it was subtracted from the original sub-image (Fig. \ref{fig2}c).

We then created annuli based on the optical ellipse parameters at increasing radii around the optical centre of the galaxy. For each annulus we measured the total flux, the surface brightness, the aperture noise (described in section \ref{aperr}) and the S/N. The S/N was calculated as the ratio of the total flux within the annulus to the noise in an aperture with the same area as the annulus.

The S/N radial profile was then used to define a cut-off once the S/N dropped below 2. This is then used to define the edge of the galaxy (the black dashed line in Figs. \ref{fig2}(d), (e), (f) \& (g)). The final measured uncertainty is then calculated from the quadrature sum of the aperture noise for the entire aperture, the calibration uncertainty and the zero-point error. The new estimate of the galaxy edge is used to create a new mask, replacing the old 1.5 $\times$ $d_{25}$ limit, and the process repeated. Once the procedure converged, aperture corrections were applied to the flux and the aperture noise measurements in accordance with \citet{ibar} and Griffin \& North (2012, in prep). 

If the total aperture S/N measured less than 3, using this method, the sub-image was optimally searched again for a point source. This consisted of convolving each sub-image with the relevant Point Spread Function (PSF) for each band and measuring the peak emission within an area defined by the convolved PSF FWHM, centred on the optical position. To measure the noise in the PSF-convolved map we plotted the histogram of the flux values and fit the negative flux values with a Gaussian function, whose width is representative of the combined instrumental and confusion noise (\citealt{marsden}, \citealt{chapin} and references therein). In practice this required some tuning, the best results being achieved when fitting a Gaussian to a reduced range of histogram bins that encompassed a significant portion of the negative tail, the peak and a small number of bins on the positive side of the peak. The width of this Gaussian was then taken as our estimate of the noise, and detections were defined as those sources above 3$\sigma$. In the case of a detection, the noise from the PSF-convolved map was summed in quadrature with the calibration uncertainty and zero-point uncertainty to provide the total error. If the source was still undetected, the 3$\sigma$ upper limit was measured from the histogram in the PSF-convolved map.

We recorded all of the radial data and intermediate masks during the process and, they will be publicly available  online\footnote{http://goldmine.mib.infn.it/} in ASCII format as well as {\it eps} figures outlining the different steps, such as that shown in Fig. \ref{fig2}.

\begin{figure}
	\includegraphics[angle=0,width=0.45\textwidth]{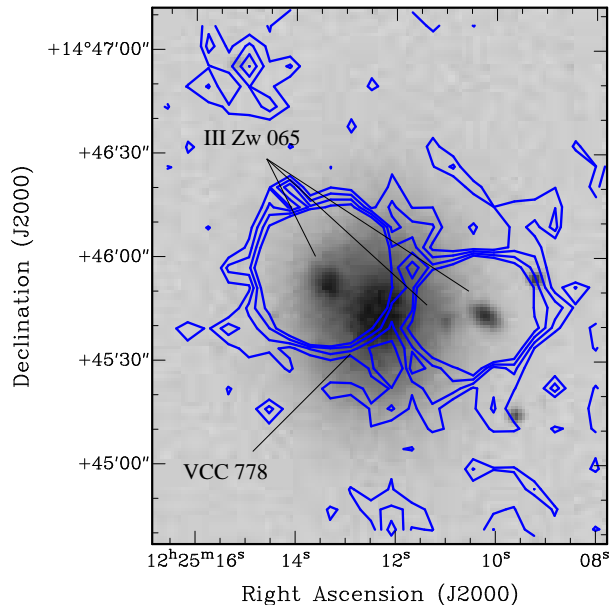}
	\caption{An example of background confusion causing an incorrect detection. A DSS greyscale image overlaid with 250 \micron\ contours at 0.005, 0.01, 0.015, 0.02 Jy \perbeam. The target early type galaxy, VCC778 is sandwiched between a background triplet (listed in {\sc ned} as III Zw 065). Their FIR emission dominate the total flux in this region, hence the emission from VCC778 is not well determined.\label{fig1a}}
\end{figure}

\begin{figure}
	\includegraphics[angle=0,width=0.45\textwidth]{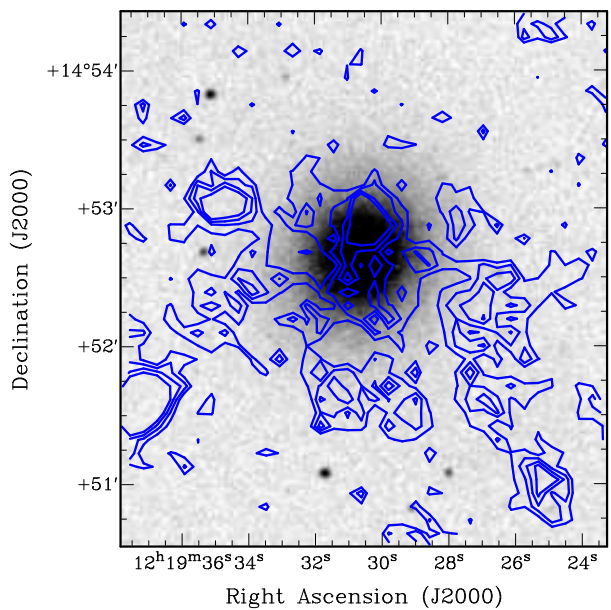}
	\caption{An example of cirrus/background contamination causing
          an ambiguous detection. A DSS greyscale image overlaid with
          250 \micron\ contours at 0.005, 0.01, 0.015, 0.02 Jy
          \perbeam. The optical centre of the target early type
          galaxy, VCC355, is offset from the peak of the 250
          \micron\ emission and the 250 \micron\ emission is in a
          region of high confusion. Since the emission cannot be
           associated with the galaxy, it has been excluded from the catalogue. \label{fig1c}}
\end{figure}

Once the automatic flux measurements were completed each detected galaxy was analysed by-eye to confirm that the masking and cut-off radius were realistic. Optical images from {\sc goldmine} and the Digital Sky Survey were examined in order to make the most comprehensive judgment for discriminating between background sources and galactic features. We found that occasionally the algorithm could be deceived by nearby, bright background galaxies or by the presence of bright cirrus which was insufficiently modelled and removed. Some were obvious as in the case of VCC778 (Fig. \ref{fig1a}); here the FIR emission consists of multiple point sources with optical background galaxies clearly visible near the centre of the emission. Some were more ambiguous as in VCC355 (Fig. \ref{fig1c}); while this galaxy does have FIR emission within the optical extent of the galaxy, it is in a highly confused region and there is nothing to distinguish it from the immediate background. For these sorts of detections, human intervention was required to define the radial cut-off manually or to exclude the galaxy from the list of detections. This affected $\sim 10$\% of the total number of measurements, most of which were false detections. In the event that the galaxy was excluded due to a contaminating source, the measured flux of the contaminating source at the position of the galaxy has been recorded as the upper limit and these sources are labelled with an asterisk in Table \ref{fluxtbl}.

Since the 250 \micron\ data has the best combination of sensitivity and resolution, we also imposed a strict requirement that a galaxy be detected at 250 \micron. This further restricts the possibility of illegitimate background sources making it into the catalogue. 

The FIR emission from the interacting galaxy pair NGC4567/8 (VCC1673/6) is a confused system with overlapping contributions from both galaxies. We have adapted the method described above to model NGC4567 and hence measure the individual fluxes. This method is fully described in Appendix \ref{ngc45678}. 

The nominal SPIRE and PACS flux calibration schemes assume a `white' point source, i.e. a compact object whose emission is flat across the filter bandpass (the K4 term). This term varies as a function of the object size and spectral index, and so must be corrected for. The fluxes presented in Table \ref{fluxtbl} use the nominal K4 values, allowing the reader to apply the corrections separately as described in the SPIRE Observer's Manual\footnote{http://herschel.esac.esa.int/Docs/SPIRE/html/spire\_om.html} and the PACS Observer's manual \footnote{http://herschel.esac.esa.int/Docs/PACS/html/pacs\_om.html}. We leave the application of the correction factors to the reader, since they require prior knowledge of the extent (which can be subjective) and colour of the source. They are also tabulated in \citet{hevics8} for a $\beta=2$ modified black-body. 

\subsection{Total uncertainty estimate}

The final uncertainty measurements quoted in Table \ref{fluxtbl} consist of contributions from calibration uncertainty, $\sigma_{cal}$, aperture uncertainty, $\sigma_{aper}$ and zero-point uncertainty, $\sigma_{zero}$. These contributions are added in quadrature thus:
\begin{equation} \label{eqn1}
\sigma_{total}^2 = \sigma_{cal}^2 + \sigma_{aper}^2 + \sigma_{zero}^2 
\end{equation}

We now discuss the derivation of each of the terms in Eq'n \ref{eqn1} before moving on to the verification of the Herschel calibration.

\subsubsection{calibration uncertainty}

For SPIRE, $\sigma_{cal}$ is based on scans of a single source (Neptune) and comparing it to a model of Neptune's emission. The final values for each band include a correlated 5\% error from the assumed model for Neptune and a ~2\% random component from repeated measurements. The Observer's manual advises that as a conservative measure, these values should be added together. $SPIRE \sigma_{cal}$ is taken to be 7\% for each band.

The situation is more complicated for PACS since the calibration is based on multiple sources with different models. The uncorrelated uncertainties of 3\% at 100 \micron\, 4\% at 160 \micron\ and a 2.2\% correlated uncertainty are given for point sources, which were observed, reduced and analysed in a different way to the HeVICS PACS data. A recent Technical Report\footnote{PICC-NHSC-TR-034} found that PACS 160 \micron\ data agreed with MIPS 160 \micron\ data to within 5-20\%, but again, this was using two different map-making schemes to that presented here. We have erred on the side of caution and assumed a value of 12\% for both PACS 100 and 160 \micron\ data, including the 2.2\% correlated error.

\subsubsection{aperture uncertainty}\label{aperr}

We followed the method of \citet{ibar} for estimating the aperture uncertainty. We recorded the total flux within random square apertures of a given size which were laid down on the entire map. The standard deviation of these totals was measured and a 3-$\sigma$ cut was then applied. The standard deviation was then re-measured and a further cut applied. This was repeated until convergence and the final value of the standard deviation was taken as the aperture noise for that sized aperture. This method invariably includes a contribution from the confusion noise in the final estimate and we make no attempt to separate the confusion noise from the instrumental noise. 

\begin{figure*}
	\includegraphics[height=0.88\textheight]{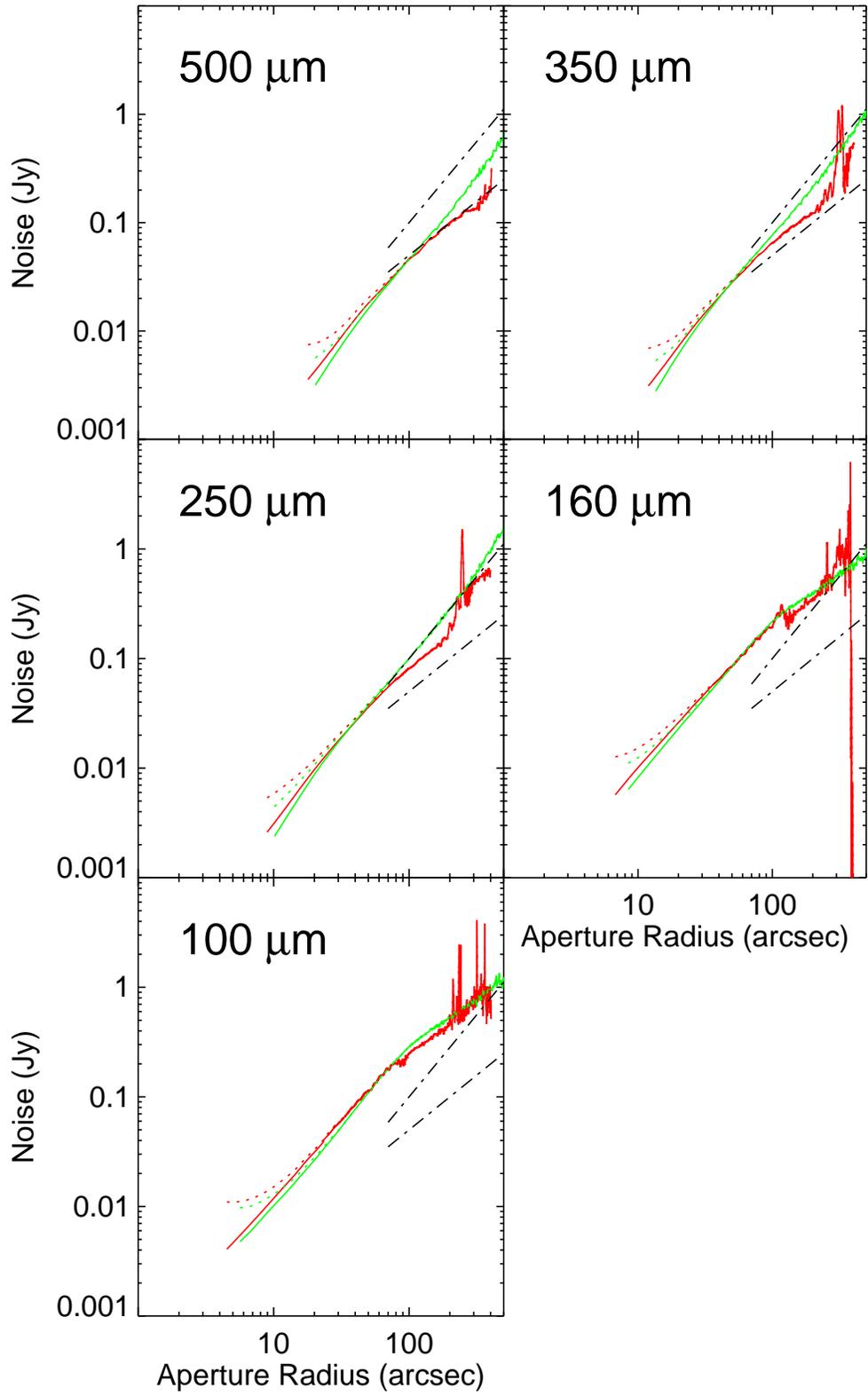}
	\caption{Variation of aperture noise with circular aperture radius. Green lines are apertures measured from the entire, full-depth region within the southern tile. Red lines show the combined results from sub-images that were extracted from the entire map to perform photometry on individual targets. Dotted lines show the same data with aperture corrections applied. Overplotted dot-dashed lines show $\sigma_{aper} \propto n_{px}^{0.5}$ and $\sigma_{aper} \propto n_{px}^{0.75}$ for comparison. \label{fig3a}}
\end{figure*}

We combined results from individual sub-images that were used to perform the photometry, and plotted the variation of aperture noise with aperture size for each band (Fig. \ref{fig3a}). In this plot the horizontal axis corresponds to the radii of circular apertures whose areas are equal to the square apertures described above. As a sanity check, we also performed the same analysis using the entire southern-most tile, since it has the least cirrus contamination. The entire tile was rebinned using pixels equal to the aperture size. A simple background value was subtracted based on the mean of the surrounding, re-binned pixels. Iterative sigma-clipping was then performed on the rebinned, background-subtracted image.

In all the bands, the application of aperture corrections is clearly important for objects comparable to the size of the beam. If we consider first the SPIRE data. One would expect the noise to increase $\propto n_{px}^{0.5}$ for white noise, but in fact the dependence is closer to $n_{px}^{0.75}$ for both the entire tile and the sub-images. At large radii, the noise measured from the sub-images exhibits a break after which the dependence is closer to the $n_{px}^{0.5}$ as expected. 

The difference between the noise measured from the entire tile and the noise measured from sub-images probably highlights the difference between the background estimation methods; the sub-images employ a high-order polynomial whereas for the entire tile, a simple mean is taken over a comparable area. Since the noise measured in the sub-images is closer to the expected value for Gaussian noise at large radii (although admittedly, the data are noisy), this is a good indication that the background subtraction is working well at removing background variation on these scales. 

The $n_{px}^{0.75}$ dependence was unexpected. At first we thought that it as probably due to the contributions from background confusion and contaminating cirrus that was poorly fit. But the same behaviour was noticed in the PACS data which has much less contamination from background sources and cirrus. Most likely, there is another source of error which is unaccounted for, perhaps from the map-making scheme itself.

\subsubsection{zero-point error}

Zero-point errors were estimated from the map of the background fit. The standard deviation of the background was measured over the extent of the source and then multiplied by the number of pixels in the source aperture. This only contributed to a significant proportion of the uncertainty for extended, low-surface brightness objects, which were rare.

\subsection{Verification of flux measurements}
The flux measurements from PACS were verified by comparing our measurements to previous measurements of Virgo galaxies from ISO 100 \& 170 \micron\ \citep{tuffs2002}, the IRAS Point Source Catalogue and the IRAS Faint Source Catalogue \citep{Moshir1993}. We applied colour corrections to the PACS fluxes in  accordance with the instrument team recommendations\footnote{http://herschel.esac.esa.int/twiki/pub/Public/PacsCalibrationWeb/cc\_report\_v1.pdf} in order to compare our measurements with those taken with ISO. When obtaining these correction factors, we assumed a blackbody temperature of 20K, typical for dust in these galaxies \citep{hevics8}. No colour correction is required to compare PACS 100 \micron\ with IRAS 100 \micron\ data. 

A cross-match of sources yielded 46 matches from \citet{tuffs2002}, 70 from the IRAS PSC and 83 from the IRAS FSC. Although instrument-independent measurements at 250, 350 and 500 \micron\  do not exist for the Virgo galaxies we have compared our results, which are based on an automatic measuring technique, with those of the BGS and the Herschel Reference Sample \citep{ciesla-hrs} which employ alternative, manual aperture photometry. For the BGS we used the original apertures but applied them to the new datasets to account for any change in flux calibration between the old data and the data presented here.

\begin{table}
  \centering
  \begin{minipage}[c]{73mm}
  \caption{Linear fit parameters to comparisons of Herschel fluxes with literature values. Uncertainties are shown in parentheses.\label{tbl3}}
  \begin{tabular}{lccc}
      \hline\\
    Wavelength & Gradient & Intercept & scatter\\          
(\micron) & & &\\
      \hline\\
100\footnote{BGS (Davies et al. 2011)} & 0.97 (0.02) & -0.03 (0.02) & 0.08\\
100\footnote{ISO (Tuffs et al. 2002)} & 1.02 (0.04) & 0.16 (0.03) & 0.13\\
100\footnote{IRAS PSC} & 1.11 (0.02)& -0.08 (0.02) & 0.05\\
100\footnote{IRAS FSC} & 1.11 (0.02) & -0.09 (0.02) & 0.08\\
&&\\
160$^a$ & 1.041 (0.006)& -0.048 (0.006) & 0.02\\
160/170$^b$ & 1.08 (0.04) & -0.01 (0.03) & 0.13\\
&&\\
250$^a$ & 1.012 (0.004) & -0.006 (0.003) & 0.02\\
250\footnote{HRS (Ciesla et al. 2012)} & 0.997 (0.004) & 0.020 (0.003) & 0.02\\
&&\\
350$^a$ & 1.015 (0.006) & -0.012 (0.003) & 0.02\\
350$^e$ & 0.989 (0.004) & 0.017 (0.003) & 0.02\\
&&\\
500$^a$ & 0.994 (0.007)& 0.009 (0.006) & 0.03\\
500$^e$& 0.98 (0.01)& 0.028 (0.005) & 0.03\\
\hline\\
  \end{tabular}
  \end{minipage}
\end{table}

\begin{figure*}
	\includegraphics[height=0.88\textheight]{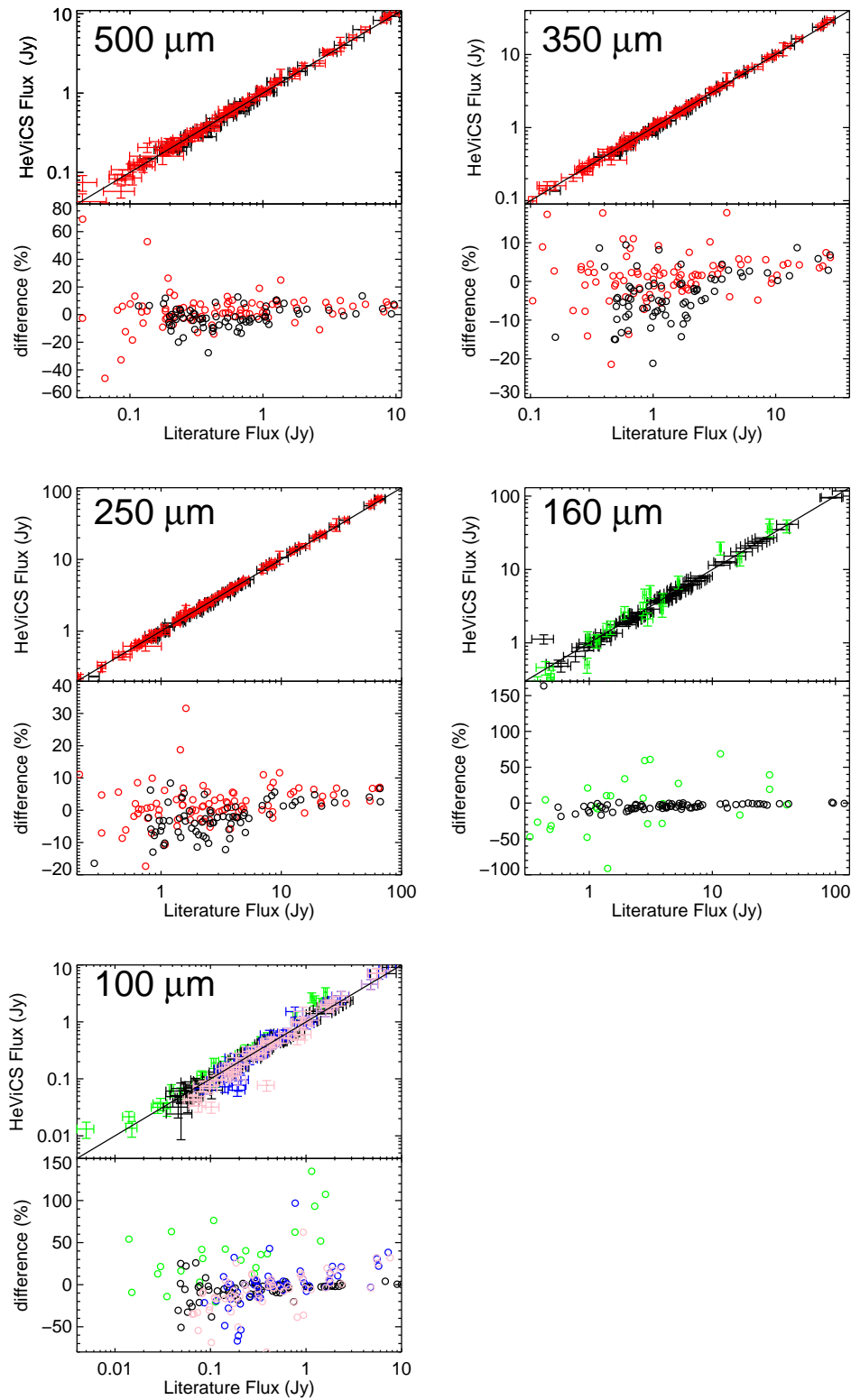}
	\caption{Herschel flux comparisons with ISO (green open circles), IRAS Point Source Catalogue (blue open circles), IRAS Faint Source Catalogue (pink open circles) , the HeViCS BGS (open black circles) and the Herschel Reference Sample (open red circles). {\it top left}: 500 \micron\ {\it top right}: 350 \micron\ {\it middle left}: 250 \micron\ {\it middle right} 160 \micron\ {\it bottom left}: 100 \micron. The 1:1 fit lines is shown in black, see text for details of individual fits to the plots. The percentage difference between HeViCS and the literature values are given below each plot. The outlier at 160 \micron\ is VCC785 (see text for further details).\label{fig3}}
\end{figure*}

Straight-line fit parameters (with 1$\sigma$ uncertainties) to the log-log plots displayed in Fig. \ref{fig3} are shown in Table \ref{tbl3}. In most cases the fit values are consistent with a 1:1 correlation, to within the scatter in the residuals, so we have high confidence in the fluxes in our catalogue for which there are no previous measurements. There is a tendency to measure higher fluxes in PACS 100 \micron\ compared to IRAS 100 \micron\ data, but this tendency has already been independently identified by the instrument team and is currently under investigation. There is a notable discrepancy between our measurements and the BGS measurements of VCC785 in the 160 \micron\ data at $\sim0.45$ Jy. The immediate area surrounding the galaxy at 160 \micron\ has significant negative flux values, and the original BGS aperture, which is based on the extent of the galaxy at 500 \micron\, envelops these features, reducing the total flux substantially.

\subsection{SED Fitting}
We have attempted to fit the SED for each galaxy under the assumption that the emission is entirely from thermal dust. M87 (VCC1316), M84 (VCC763) and NGC4261 (VCC345) were omitted since these galaxies have significant synchrotron emission that could potentially contaminate the SPIRE fluxes (\citealt{hevics6}, \citealt{smith-etg}). We employed a single-temperature modified blackbody model with a fixed emissivity index, $\beta=2$, to fit all the galaxies for which we had flux measurements across all bands. The dust mass-opacity coefficient used was 0.192 m$^2$ kg$^{-1}$ at 350 micron \citep{draine}. We realise that adopting a single-temperature fit and a fixed $\beta=2$ may be viewed as contentious in the light of recent results (e.g. \citealt{planck-smclmc}, \citealt{dale}, \citealt{boselli-hrs}) however the BGS galaxies have been shown to be well modelled by a single-temperature, $\beta = 2$ fit (\citealt{hevics8}, \citealt{hevics9}). $\beta$ is still a poorly understood quantity and may not even be constant within a single galaxy (\citealt{planck-smclmc}, \citealt{smith-m31}). We have chosen the simplest approach which is probably the best we can achieve with the wavelength coverage at our disposal and will provide a benchmark for future studies, which will exploit ancillary datasets to look at the effects of adopting multi-temperature, variable $\beta$ models. We also acknowledge that the range of $\beta$ that exist in the literature can typically lead to a factor of a few difference in the derived dust mass. This can rise to a factor of 10 if an extreme $\beta =1$ model is assumed \citep{cortese-ltg}.

The SED-fitting algorithm works by minimising the difference between the set of measured fluxes and a set of model fluxes at the same wavelengths. The model fluxes are generated by taking a modified blackbody function and multiplying it by the Herschel filter profiles. This step requires measured fluxes that have had the original K4 scaling factor removed. A simplex technique is used to find the most likely combination of temperature and mass and several thousand Monte-Carlo simulations are then employed to derive the error distributions for the dust mass and temperature. These simulations lend a greater degree of robustness over simpler $\chi^2$-minimising algorithms which might assume that the error distributions are Gaussian. The method also takes into account uncorrelated calibration uncertainties for SPIRE and PACS (see section \ref{dr} for these values) and correlated calibration errors between SPIRE bands which are of the order of 5\%. The results of the Monte-Carlo simulations indicated that the errors in dust mass and temperature are anti-correlated, but are centred on the correct values (see \citealt{smith-m31}).

In order to estimate the dust mass, it was necessary to provide an estimate for the distance. We have adopted the distances provided by {\sc goldmine} which quote the average distance to the sub-cluster to which each individual galaxy belongs, as measured in \citet{gavazzi-1999}.

In some cases it was clear, both from the $\chi^2$ values and from visual inspection of the fits that a single temperature fit was an inappropriate choice. We have not attempted further modelling of these galaxies since future papers will examine multi-temperature models, employing variable $\beta$, with the help of additional datasets which will provide the necessary extra wavelength coverage to better constrain the more complex models (Grossi et al. 2012, in prep.).

\section{Results}
\label{results}
\subsection{Detection Rates}
\begin{table}
  \centering
  \caption{Herschel detection rates of VCC galaxies contained within the HeViCS survey limits. Results are given for the whole survey region and the inner region covered to full depth. \label{tbl2}} 
  \begin{tabular}{lcc}
\hline\\
Wavelength & Entire Survey region & 8-scan Survey region \\
(\micron) & (84 sq. deg.) & (inner 55 sq. deg.)\\
&(750 objects)&(602 objects)\\
\hline\\
500 & 204 (27\%) & 160 (27\%)\\
350 & 232 (31\%) & 184 (31\%)\\
250 & 254 (34\%) & 203 (34\%)\\
160 & 210 (28\%) & 176 (29\%)\\
100 & 195 (26\%) & 165 (27\%)\\
\hline\\
  \end{tabular}
\end{table}

\begin{figure*}
\centering
\subfigure[Distribution of FIR detections sorted by photographic apparent magnitude, $m_{pg}$.]{
	\includegraphics[height=0.4\textheight]{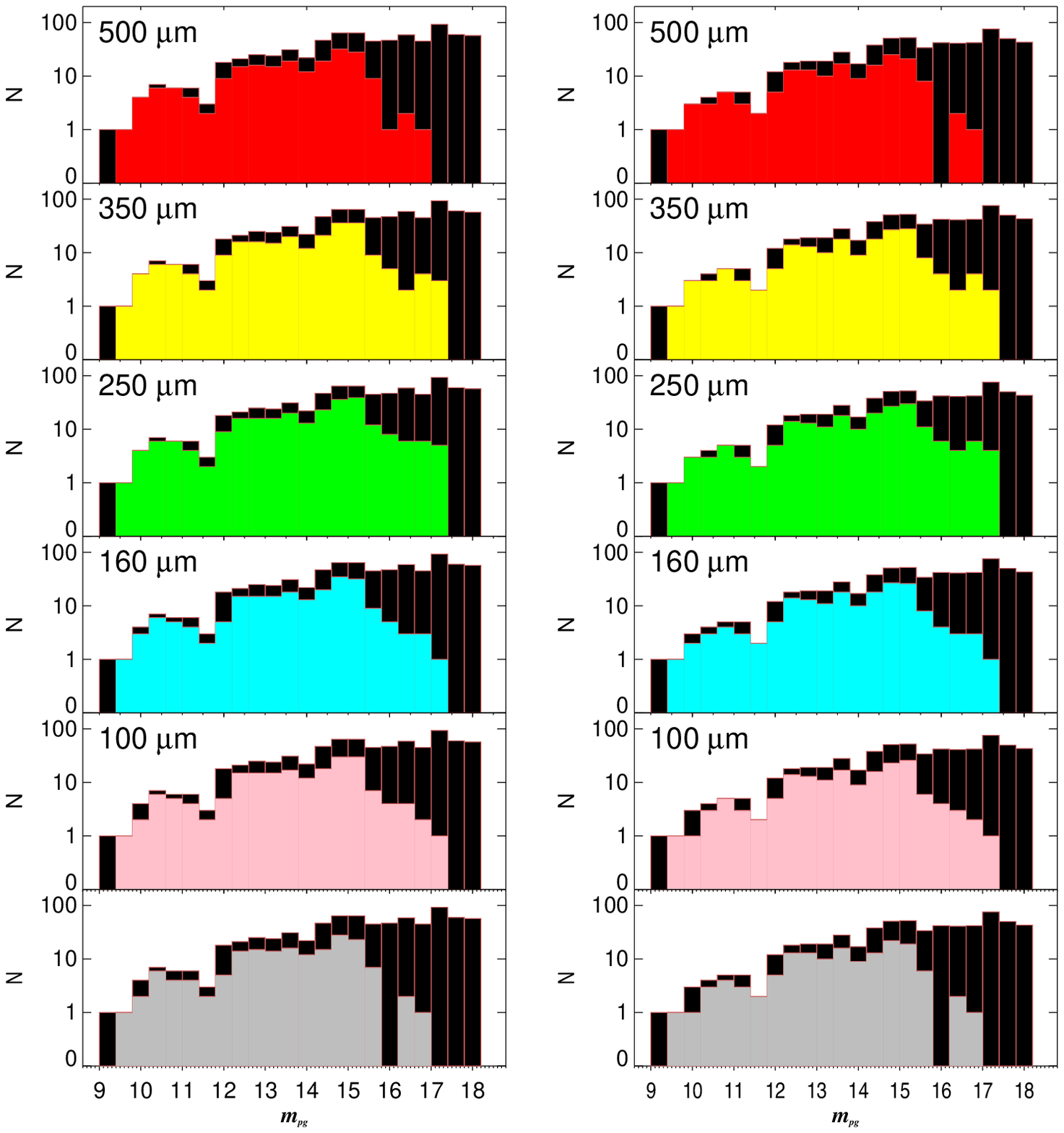}
	\label{fig4a}}
\subfigure[FIR detections sorted by morphological type.]{
	\includegraphics[height=0.43\textheight]{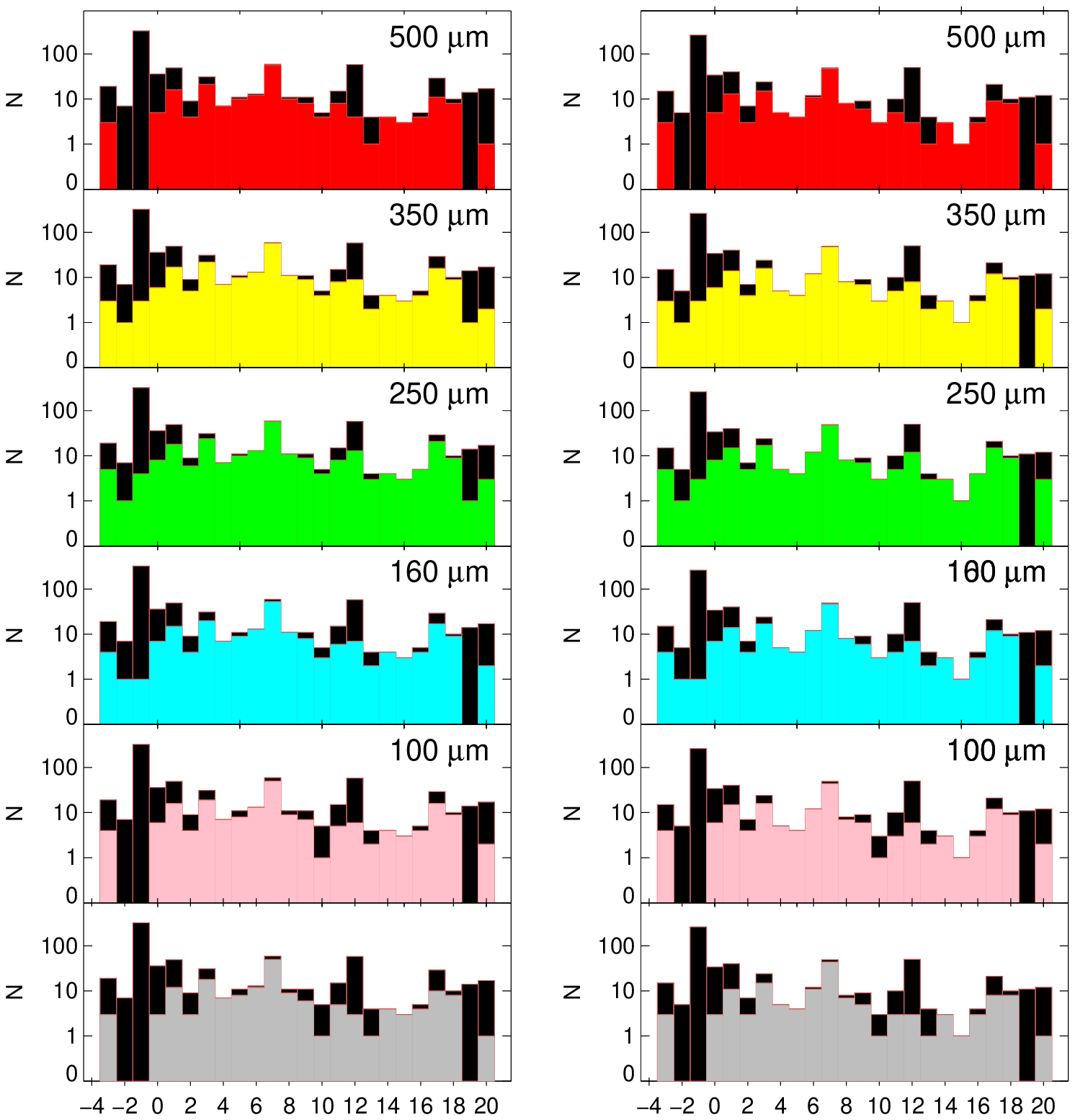}
	\label{fig4b}}
	\caption{FIR detections for the VCC galaxies. The left hand panels correspond to the entire survey region (84 sq. deg.), the right hand panels to the central 55 sq. deg. with optimum coverage. Black represents the optical detections of the VCC galaxies down to the completeness limit, $m_{pg} \leq 18$. 500\micron, 350\micron, 250\micron, 160\micron, 100\micron\ \& 5-band detections are represented by red, yellow, green, cyan, pink and grey histograms respectively. Galaxy morphology is as follows: {\bf -3}: dS0, {\bf -2}: dE/dS0, {\bf -1}: dE (d:E), {\bf 0}: E - E/S0, {\bf 1}: S0, {\bf 2}: S0a - S0/Sa, {\bf 3}: Sa, {\bf 4}: Sab, {\bf 5}: Sb, {\bf 6}: Sbc, {\bf 7}: Sc (dSc), {\bf 8}: Scd, {\bf 9}: Sd, {\bf 10}: Sdm - Sd/Sm, {\bf 11}: Sm, {\bf 12}: Im (Im/S), {\bf 13}: Pec, {\bf 14}: S/BCD (dS/BCD dS0/BCD Sd/BCD), {\bf 15}: Sm/BCD, {\bf 16}: Im/BCD, {\bf 17}: BCD, {\bf 18}: S (dS), {\bf 19}: dIm/dE, {\bf 20}: ? \label{fig4}}
\end{figure*}

Table \ref{tbl2} displays the detection rates for the entire ~84 sq. deg. region and the inner ~55 sq. deg. covered to optimum depth. In this table we have included all galaxies that yielded a detection at 250 \micron. The breakdown of the FIR detections by $m_{pg}$ for each wavelength are shown in Fig. \ref{fig4} (top row) for the two survey regions. Surprisingly there does not appear to be any significant decrease in detection rates in the regions with less coverage. This is probably suggesting that Herschel is sensitive enough to detect all the FIR-emitting Virgo galaxies within a couple of passes. Individual fluxes for each galaxy can be found in Table \ref{fluxtbl}.

\begin{figure}
\centering
	\includegraphics[width=0.48\textwidth]{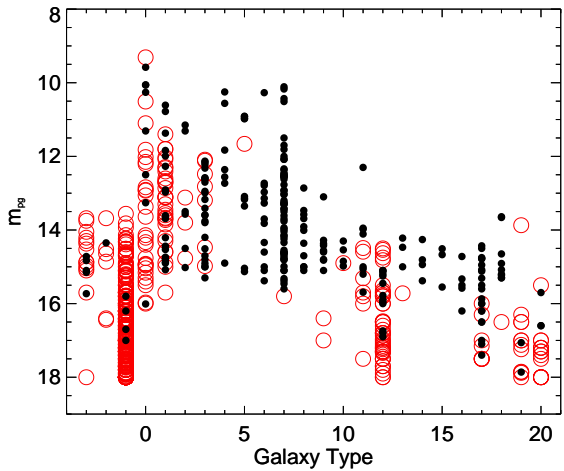}
	\caption{Morphology-magnitude plot for the VCC sample within the entire HeViCS area. Galaxies with a detection in at least one band are black solid circles, while undetected galaxies are open red circles. Galaxy morphology numbering follows that in Fig. \protect{\ref{fig4}}. \label{fig5a}}
\end{figure}

The difference between FIR detections (predominantly thermal dust) and optical detections is evident at low optical luminosities in Figs. \ref{fig4} and \ref{fig5a}. The galaxy cluster population is dominated by low luminosity dwarfs, in particular dwarf ellipticals (dEs), but few of these galaxies are detected in FIR. Since these galaxies are deficient in dust, it is unsurprising that their detection rates are low. This is also demonstrated in Fig. \ref{fig4b}, which displays the same information but broken down by galaxy type. The other early-types also have low detection rates, as these too have little dust present in the ISM. The late-types have high detection rates in both PACS and SPIRE as one would expect since these galaxies are rich in dust.

Fig. \ref{fig5a} shows the morphology-magnitude plot, highlighting detected and undetected galaxies. This plot allows us to compare different galaxy morphologies at the same optical magnitude. From this it is clear that the lack of FIR detections of dEs, ellipticals (Es) and irregulars (Im/S) is not simply because they are faint at optical magnitudes - there are many FIR non-detections which have an optical magnitude greater than the faintest detection.

The Magellanic-type dwarfs are notable by their lack of detection rates amongst the late-types. Although these systems are metal-poor and would not necessarily be expected to contain much dust, they are of similar metallicity to the blue compact dwarfs (BCDs), which have a higher detection rate. This suggests that the lack of Sm/Im galaxies is more than likely due to these galaxies falling below the surface brightness detection threshold.

Although the early-types are more readily detected in the optical compared to the FIR, the bins are by no means empty as one might expect of a population of dust-deficient systems. These exceptional galaxies have been the focus of previous HeViCS studies (e.g. \citealt{hevics5}, \citealt{hevics7}, \citealt{smith-etg}) and will be the subject of future analyses as these new data are fully exploited (di Serego Alighieri, 2012 in prep. Grossi et al. 2012, in prep.).

\subsection{Background Contamination}
As explained in Section \ref{fluxmeas}, during the formation of the catalogue, some FIR detections were rejected either because the FIR emission did not coincide with a foreground optical galaxy, or could not be distinguished from the surrounding cirrus. Despite the precautions taken to avoid contamination from background sources, there is still a possibility of accidentally reporting the flux from a background galaxy. 

In order to estimate this effect, we have adopted a na\"ive approach. We restricted the analysis to the total population of point sources and undetected sources. Extended sources at the expected positions of Virgo galaxies can be reliably attributed to those galaxies. We then assumed that the background galaxy population is randomly distributed on the sky and worked out the probability of a chance alignment within each SPIRE beam, based on the source counts of \citet{glenn}. This probability is then used to predict the number of possible spurious sources in the catalogue in different flux bins.

If one has chosen the rejected galaxies wisely, then one would expect that the number of spurious measurements should be approximately equal to the predicted number of contaminating background sources. In Table \ref{conttbl} we show the results of this test. From this table we see that at 500 \micron\ the number of rejections is approximately equal to the number of predicted sources. At 350 \& 250 \micron\ we appear to have been overzealous in excising galaxies at the lowest fluxes. We attribute at least some of the increase in the number of spurious detections to peaks in the cirrus, which are expected at these low fluxes. Above 45 mJy the number of rejected sources compares well with the predictions. 

The agreement between the number of spurious sources and the number of predicted background sources does not guarantee that the correct set of galaxies has been selected in the catalogue. However, the fact that the selection was made by comparing FIR emission with optical data does provide a greater degree of reliability. The combination of these points suggests it is likely that none of the catalogue has been misidentified.

\begin{table}
  \centering
  \begin{tabular}{lccccc}
\hline\\
Flux bin & $N_{a}$ & $N_{r}$ & $N_{p}$ & contamination & $N_{c}$\\
(mJy) & & & (deg$^{-2}$) & \% & \\
\hline\\
{\bf 250 \micron} & & & & & \\
15 -- 20 & 8 & 47 & 1694 & 5.5 & 28 \\
20 -- 45 & 20 & 41 & 1824 & 5.9 & 31\\
45 -- 100 & 6 & 6 & 313 & 1 & 5 \\
100+ & 4 & 1 & 25 & 0.08 & 0 \\
\\
{\bf 350 \micron} &&&&&\\
17 -- 20 & 1 & 27 & 566 & 3 & 17\\
20 -- 45 & 20 & 55 & 1209 & 7 & 38\\
45 -- 100 & 6 & 4 & 154 & 0.9 & 5 \\
100+ & 4 & 1 & 73 & 0.3 & 1.6\\
\\
{\bf 500 \micron} &&&&&\\
17 -- 20 & 3 & 14 & 185 & 2 & 11\\
20 -- 45 & 29 & 45 & 614 & 7 & 41\\
45 -- 100 & 3 & 2 & 19 & 0.2 & 1\\
100+ & 2 & 0 & 1 & 0.01 & 0\\
\hline\\

\end{tabular}
\caption{Estimates of contamination from background galaxies in the SPIRE bands.  $N_{a}$ is the number of sources that were accepted into the catalogue, $N_{r}$ is the number of sources that were rejected as spurious, $N_{p}$ is the number density of sources predicted from the number counts of \protect{\citet{glenn}}. Contamination is the expected percentage contamination based on the source counts. $N_{c}$ is the expected number of contaminating sources.\label{conttbl}}
\end{table}

\subsection{FIR colours}
\begin{figure*}
\subfigure{
	\includegraphics[height=0.44\textheight]{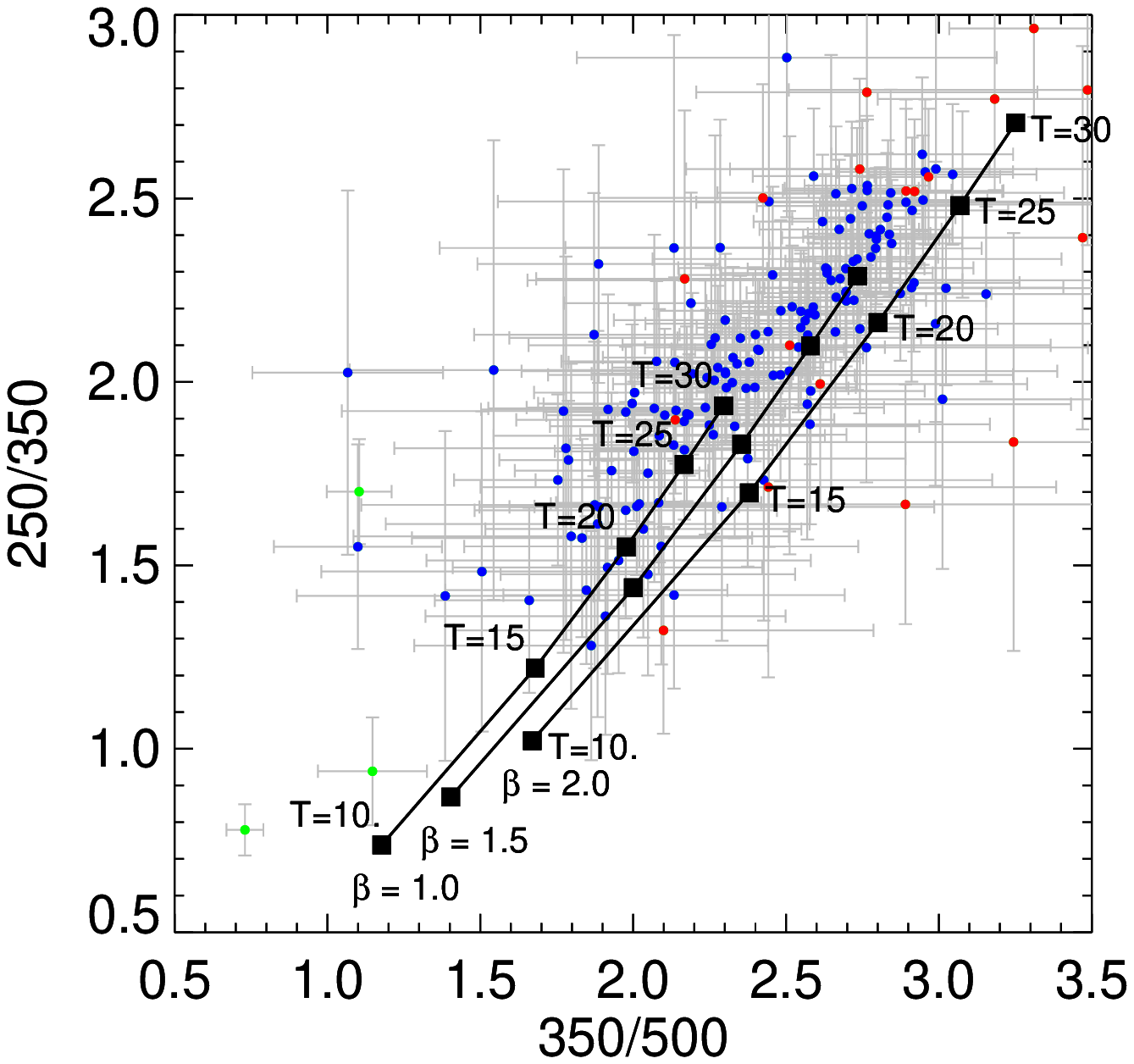}
	\label{fig3b}}
\subfigure{
	\includegraphics[height=0.44\textheight]{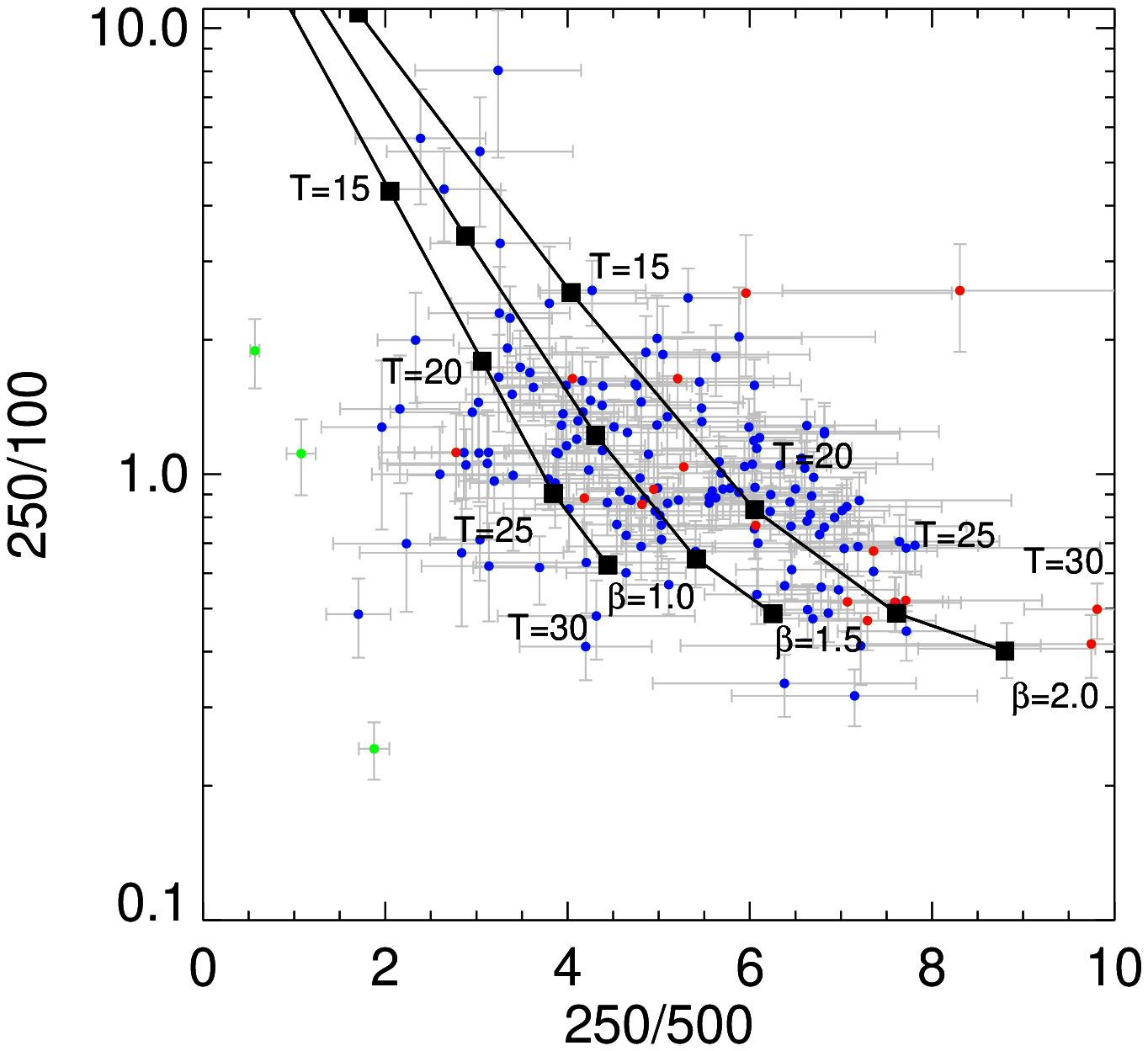}
\label{fig3c}}
\caption{Colour-colour plots for the entire HeViCS catalogue detected in all bands. Blue points are late-types, red points are early-types and green points are galaxies known to harbour AGN. Loci connecting models of constant $\beta$, but variable temperature are overplotted in black, with different temperatures labelled every 5 K.}
\end{figure*}

Recently, \citet{boselli-hrs} examined SPIRE colours for the HRS sample of 322 nearby galaxies. They observed a number of trends between FIR colour, metallicity, star formation history, the ionising radiation field and surface brightness. They also used the SPIRE only colours to argue that when fitting a fixed emissivity index, single-temperature modified blackbody model to the FIR SED, $\beta=2$, is inappropriate for all but the most metal-rich, high surface brightness galaxies in the HRS. However, due to the degeneracy between temperature and $\beta$, SPIRE-only colours are not adequate to constrain $\beta$. We demonstrate this with Figures \ref{fig3b} \& \ref{fig3c} which compare SPIRE only colours with SPIRE-PACS colours.

The fluxes that were used to create these figures were derived using the fluxes from Table \ref{fluxtbl}, and dividing out the $K_4$ factor. This provides us with a set of instrumental fluxes, that have been corrected for the size of the aperture, with no assumption on spectral index. We then generated a set of ideal modified blackbody functions with varying $\beta$ and temperatures. These models were multiplied by the instrument filter functions to reproduce the instrumental fluxes for each band. The black filled squares in Figs \ref{fig3b} \& \ref{fig3c} represent the theoretical colours derived from these model instrumental fluxes. Loci joining models of constant $\beta$ but variable temperature were then overplotted (black lines).

Grey bars indicate the uncertainty in each colour. These have been derived by adding the relative flux uncertainties in quadrature for a given colour. Since colours common to one instrument are unaffected by correlated uncertainties, we have removed the 5\% correlated error, when deriving the the SPIRE, $S250/S500$ colour uncertainty.

A comparison of Fig. \ref{fig3b} with \citet{boselli-hrs} Fig. 9 shows that the galaxies in HeViCS exhibit similar trends in SPIRE colours to those in the HRS.  There appears to be two clusters of objects around the $\beta=1.5$ line and $\beta=1$, but the uncertainties in the colours prevent one from disentangling the temperature-$\beta$  degeneracy. So the cluster of galaxies around ($\beta=1, T=30$ K), could also be consistent with a ($\beta=2, T=17$ K) model.

The situation improves when a PACS colour is introduced (Fig. \ref{fig3c}). It is easier to distinguish between different models of $\beta$ and $T$, and we are able to constrain the temperature with much greater accuracy. We now see that the HeViCS galaxies exhibit a range of $\beta$ from 1 -- 2 and the cold ($T < 15$K) galaxies are unambiguously cold. Most of the galaxies appear to cluster around the models corresponding to $\beta = 1.5$ \& $\beta=2$ with temperatures varying between 20 -- 30 K. This is very similar to the $\beta, T$ distributions observed by \citet{duneales-slugs-2001} and typical of values used in the literature (\citealt{magnelli}, \citealt{skibba}, \citealt{smith-etg}, \citealt{dungomez-2011}, \citealt{roseboom}, \citealt{hevics8}). 

There is also a cluster of objects that are consistent with $\beta = 1$ models. Values of $\beta$ as low as one are often observed in individual regions within the Milky Way (\citealt{dupac}, \citealt{bracco}) but are rarely seen in the SEDs of galaxies which are integrated over the whole galaxy. This region of the colour-colour plot becomes increasingly sensitive to contributions from either synchrotron or free-free emission (which would boost the 500 \micron\ relative to the 250 \micron), and warmer dust (which would boost the 100 \micron\ relative to the 250 \micron). The known AGN have the reddest colours but surprisingly we find a late-type, VCC223, in amongst them. An analysis of the SED reveals that it has a high 500 \micron\ flux relative to the other bands. This excess could be due to free-free or synchrotron emission from recent star formation as this galaxy is classed as a BCD. However, we can't rule out the possibility that this galaxy harbours a very cold dust component.

With such a large spread in the values of $\beta$ implied by the data, it is a useful exercise to explore the validity of a single-temperature, fixed-beta modified blackbody model, which are commonly used throughout the literature.

\subsection{SED fits, dust masses and temperatures}

\begin{figure}
\begin{center}
\includegraphics[width=0.48\textwidth]{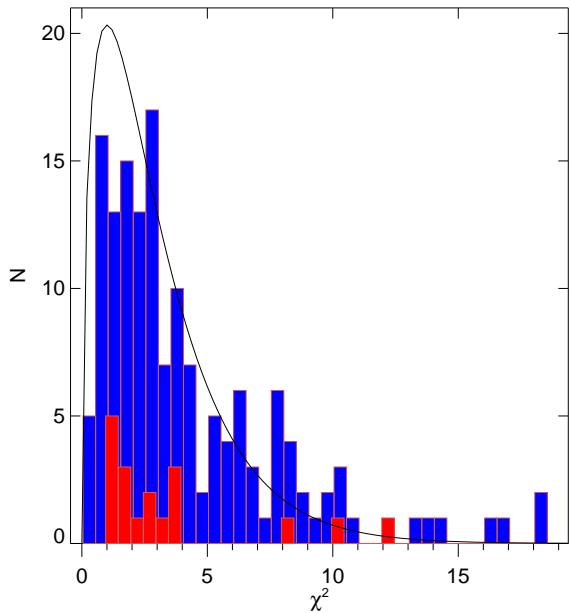}
	\caption{$\chi^2$ distribution from single-temperature, modified blackbody model fits for 161 HeViCS galaxies with 5-band photometry. Early-type galaxies are shown in red, late-types in blue. Overplotted is the theoretical $\chi^2$ distribution for 3 degrees of freedom, normalised to the bin value at the peak of the distribution. \label{fig5b}}
\end{center}
\end{figure}

We restricted our analysis of the SEDs to 168 of the 171 galaxies which were detected in all five bands of PACS and SPIRE. M87, M84 and NGC4261 were omitted since their SEDs contain significant synchrotron emission and they, along with the other early-type VCC galaxies, will be the focus of a forthcoming paper (di Serego Alighieri et al. 2012 in prep.). The distribution of $\chi^2$ for the 168 galaxies is shown in Fig. \ref{fig5b}. For 140 galaxies the single-temperature model produced an adequate fit ($\chi^2_{dof=3} < 7.8$, at a confidence level of 95\%) so that no further fitting was required. For the remaining 28 galaxies, the single-temperature model was unsuitable. This number of galaxies is significantly larger than the $\sim8$ (5\%) expected from the $\chi^2$ distribution for 3 degrees of freedom, providing quantitative evidence that the single-temperature, $\beta=2$  model is not appropriate for the FIR SEDs of all of these galaxies. Modelling of different dust populations will be studied in a future paper.

Out of the 429 early-type galaxies in the VCC that are located inside the HeViCS field, only 15 meet the requirement of $\chi^2_{dof=3} < 7.8$ . This should be kept in mind in the following simple analysis since these galaxies are clearly unusual for their type. Dust masses and temperatures are recorded in Table \ref{temptbl}. On inspection, the upper and lower limits for the dust mass and temperature were symmetric to 1 significant figure, so we have just quoted one value. The SEDs are shown in Fig. \ref{fig6a}. The $\chi^2$ for each fit is recorded at the top of each plot. 

\begin{figure}
\begin{center}
\includegraphics[width=0.48\textwidth]{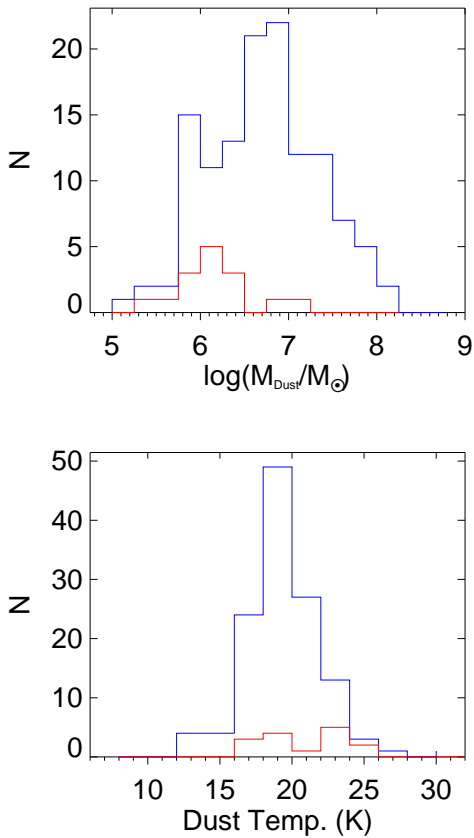}
	\caption{Temperature and mass distributions split by galaxy type. Blue columns represent types Sa and later, red represents types earlier than Sa. Only galaxies whose model fits have $\chi^2 < 7.8$ are included.\label{fig5}}
\end{center}
\end{figure}

In Fig. \ref{fig5} we show the distributions of dust mass and temperature that were derived from our SED fits, using only the 140 galaxies with a $\chi^2 < 7.8$  from the single-temperature modified blackbody spectrum model fitting. Each plot shows the different distributions based on galaxy type; types earlier than Sa are shown in red outline, while types Sa and later are shown in blue. 

The late-types exhibit a range in dust mass of log$(M_{dust}/M_{\sun}) =$5.2--8.1, with a mean value of log$(\langle M_{dust}\rangle/M_{\sun}) = 7.1 \pm 0.1$. In contrast, the early-types exhibit a narrower range of masses; log$(M_{dust}/M_{\sun}) = $5.4--7.0 , with a lower mean value of log$(\langle M_{dust}\rangle/M_{\sun}) = 6.3 \pm 0.3$. For the temperature distributions, the late-types range from 12.9 -- 26.4 K, with a mean value of $\langle T\rangle = 19.4 \pm 0.2$ K. The early-types cover a range in temperature from 16.2 -- 25.9 K, with a mean value, $\langle T \rangle = 21.1 \pm 0.8$ K. A comparison of the dust properties of the late-types is consistent with those from the BGS. This isn't surprising, since the BGS consisted of the brightest FIR/sub-mm galaxies in HeViCS, which are mostly late-type galaxies.

We performed a Kolmogorov-Smirnov (K-S) two-sample test, Mann-Whitney {\it U}-test and an F-test to look for significant differences between the two sample populations and in addition, compared the two samples to their equivalent samples in the HRS (\citealt{cortese-ltg},\citealt{smith-etg}). We note here that the HeViCS sample probes lower mass galaxies than those in the HRS, which may slightly bias the results. The late-types comparison is limited to the mass distribution since \citet{cortese-ltg} did not attempt to estimate dust temperatures for their sample.

The results of the tests are displayed in Table \ref{tbl2a}, and show that there is a clear distinction between the dust masses and temperatures of cluster late-types and cluster early-types: Virgo late-type galaxies have typically cooler, and more massive dust reservoirs than Virgo early-types. The tests also imply that late-types in the field have a significantly different mass distribution to those in the cluster. The effect of the cluster environment on the bulk dust properties of the early-types is much less pronounced, with the tests revealing a fairly high probability that the cluster and field samples were drawn from the same distribution. Since the early types are dominated by S0 galaxies in both samples, this is suggesting that the bulk dust properties of S0s are the same inside and outside the cluster. However, as noted in \citet{cortese-ltg}, differences in the dust properties may be more noticeable if the samples are defined on a property that is more indicative of a galaxy undergoing an interaction with the cluster, (e.g. HI-deficiency) rather than cluster membership alone.

\begin{table*}
  \centering
  \caption{Virgo cluster and field comparisons of the dust mass and temperature distributions of early- and late-type galaxies \label{tbl2a}}
  \begin{tabular}{cccccccccccc}
\hline\\
\multicolumn{2}{c}{Comparison samples} & $\mu_{1} $ & $\sigma_{1}$ & $\mu_{2}$ & $\sigma_{2}$& \multicolumn{2}{c}{K-S test} & \multicolumn{2}{c}{MW {\it U}-test} & \multicolumn{2}{c}{{\it F}-test} \\
1 & 2 & & & & & Value & $ p_{value}$ & Value & $p_{value}$ & Value & $p_{value}$ \\
\hline
\multicolumn{3 }{l}{{\bf Dust Mass  (log$(M_{dust} / M_{\sun})$)}} & &&&&&&\\
Cluster late & Cluster early & 7.1 & 1.6 & 6.3 & 1.2 & 0.52 & 0.0007 & -3.48 & 0.0002 & 64.27 & $10^{-10}$\\
Cluster late & Field late & 7.1 & 1.6 & 7.3 & 1.6 & 0.33 & 9$\times10^{-7}$ & 4.82 & 7$\times 10^{-7}$ & 2.58 & $10^{-7}$  \\
Cluster early & Field early & 6.3 & 1.2 & 6.6 & 1.2 & 0.29 & 0.45 & 0.67 & 0.25 & 2.61 & 0.08 \\
\multicolumn{3}{l}{{\bf Dust Temperature (K)}} & &&&&&&\\
Cluster late & Cluster early & 19.4 & 2.5 & 21.1 & 3.2 & 0.34 & 0.06 & 1.82 & 0.035 & 1.7 & 0.14\\
Cluster early & Field early & 21.1 & 3.2 & 23.5 & 4.6 & 0.31 & 0.37 & 1.50 & 0.07 & 2.06 & 0.18\\
\hline\\
  \end{tabular}
\end{table*}

One aspect of recent Herschel galactic studies that has provoked keen interest in the wider community is the existence of galaxies that seem to possess an excess of emission at 500 \micron\  (e.g. \citealt{dale}, \citealt{galametz}, \citealt{hevics5}) so it would seem pertinent to comment on their presence in the catalogue. Such galaxies are interesting because there is still disagreement on the origin of the excess. The excess could be due to an as yet undiscovered very cold dust component, thermal free-free emission, synchrotron emission or something more exotic (e.g. spinning dust \citealt{bot2010}).

We have defined the excess as those galaxies whose 500 \micron\  flux is at least  2$\sigma$ above the model fit. This yields six candidates (VCC120, VCC223, VCC562, VCC848, VCC971 \& VCC1673) with an excess at 500 \micron. VCC1673 is discounted because of the contamination from VCC1676. VCC971 (NGC4423) is an Sd; VCC223, VCC562 \& VCC848 are BCDs and VCC120 is classed as an Scd. We prefer not to speculate, at this point, as to the source of this excess, merely to highlight these objects as interesting. This is partly because six galaxies is about the amount of galaxies expected to be 2-$\sigma$ outliers, but even if the deviations have non-statistical origins, we don't have sufficient wavelength coverage in this dataset to be able to constrain more complex SEDs which might account for the excess. 

There is currently a HeViCS observing program with PACS at 70 \micron\  which should help constrain the SED at the shorter wavelengths for a substantial sample of the Virgo galaxies displayed here. In addition to this, the recent work of \citet{bendo} will provide Spitzer-MIPS photometry (24, 70 \& 160 \micron) for a significant fraction of the VCC galaxies. The combination of these two datasets will constrain the warmer ($>30$ K) dust component, while ongoing millimetre observations and archived radio observations of the emerging population showing 500 \micron\  excess, should be able to provide the necessary spectral coverage to constrain the longer wavelengths and ascertain the origin of the excess.

\section{Summary}
\label{summary}
We have undertaken the largest and deepest survey of the Virgo Cluster at FIR wavelengths with the Herschel Space Observatory, utilising both PACS and SPIRE instruments. The survey, which is now complete in 5 FIR bands, covers 84 sq. deg. of the Virgo Cluster encompassing 750 VCC galaxies brighter than the optical completeness limit. We recover 254 (34\%) of the VCC population in at least one wavelength and have 5-band FIR photometry for 171 galaxies. The photometry compares well with previous surveys showing typically less than a 10\% variation across all bands.

FIR colours from PACS and SPIRE have been compared to model single-temperature modified blackbody spectra incorporating a fixed emissivity index. The distribution of colours is consistent with a range of $\beta$=1--2. This is supported by a study of the $\chi^2$ distribution from $\beta$=2 model fits to the galaxies with 5-band photometry. The amount of galaxies not well fit by the model, is too large to be explained by random fluctuations. Further modelling is required to explore variable beta and/or different dust temperature populations, although this will likely require more measurements at FIR/MIR and submm wavelengths to be properly constrained.

A preliminary analysis of different morphological types indicates that dust in Virgo early-types has the same temperature and mass distribution as early-types in the field, suggesting that the environment has little impact on their bulk dust properties. Since the early-types are dominated by S0s, it suggests that the processes responsible for creating S0s has taken place before they enter the cluster. In contrast, the late-types have significantly different mass and temperatures to those in the field.

As well as providing a table of flux measurements and images of FIR SEDs with this manuscript, we will be providing the community with a legacy product which consists not only of the final 84 sq. deg. maps but also detailed data products such as cut-outs of the fully-processed images for each galaxy, the intermediate products that were used in the measurement process and the resulting radial profiles in each band. These will ensure the legacy value of the Herschel Virgo Cluster Survey.

\section*{Acknowledgments}
We would like to thank the anonymous referee for his/her helpful comments and  suggestions for improving this manuscript.

{\it SPIRE} has been developed by a consortium of institutes led by Cardiff University (UK) and including Univ. Lethbridge (Canada); NAOC (China); CEA, LAM (France); IFSI, Univ. Padua (Italy); IAC (Spain); Stockholm Observatory (Sweden); Imperial College London, RAL, UCL-MSSL, UKATC, Univ. Sussex (UK); and Caltech, JPL, NHSC, Univ. Colorado (USA). This development has been supported by national funding agencies: CSA (Canada); NAOC (China); CEA, CNES, CNRS (France); ASI (Italy); MCINN (Spain); SNSB (Sweden); STFC (UK); and NASA (USA). HIPE is a joint development (are joint developments) by the Herschel Science Ground Segment Consortium, consisting of ESA, the NASA Herschel Science Center, and the HIFI, PACS and SPIRE consortia.

This research has made use of the GOLDMine Database, operated by the Universit\`a degli Studi di Milano-Bicocca

The research leading to these results has received funding from the European Community's Seventh Framework Programme (/FP7/2007--2013/) under grant agreement No. 229517.

\label{lastpage}

\bibliographystyle{mn2e}
\bibliography{hevics-vcc-v2}
\onecolumn

\appendix
\section{Modelling the galaxy pair NGC4567/8 (VCC1673/6)}
\label{ngc45678}
\begin{figure*}
\begin{center}
\includegraphics[width=0.9\textwidth]{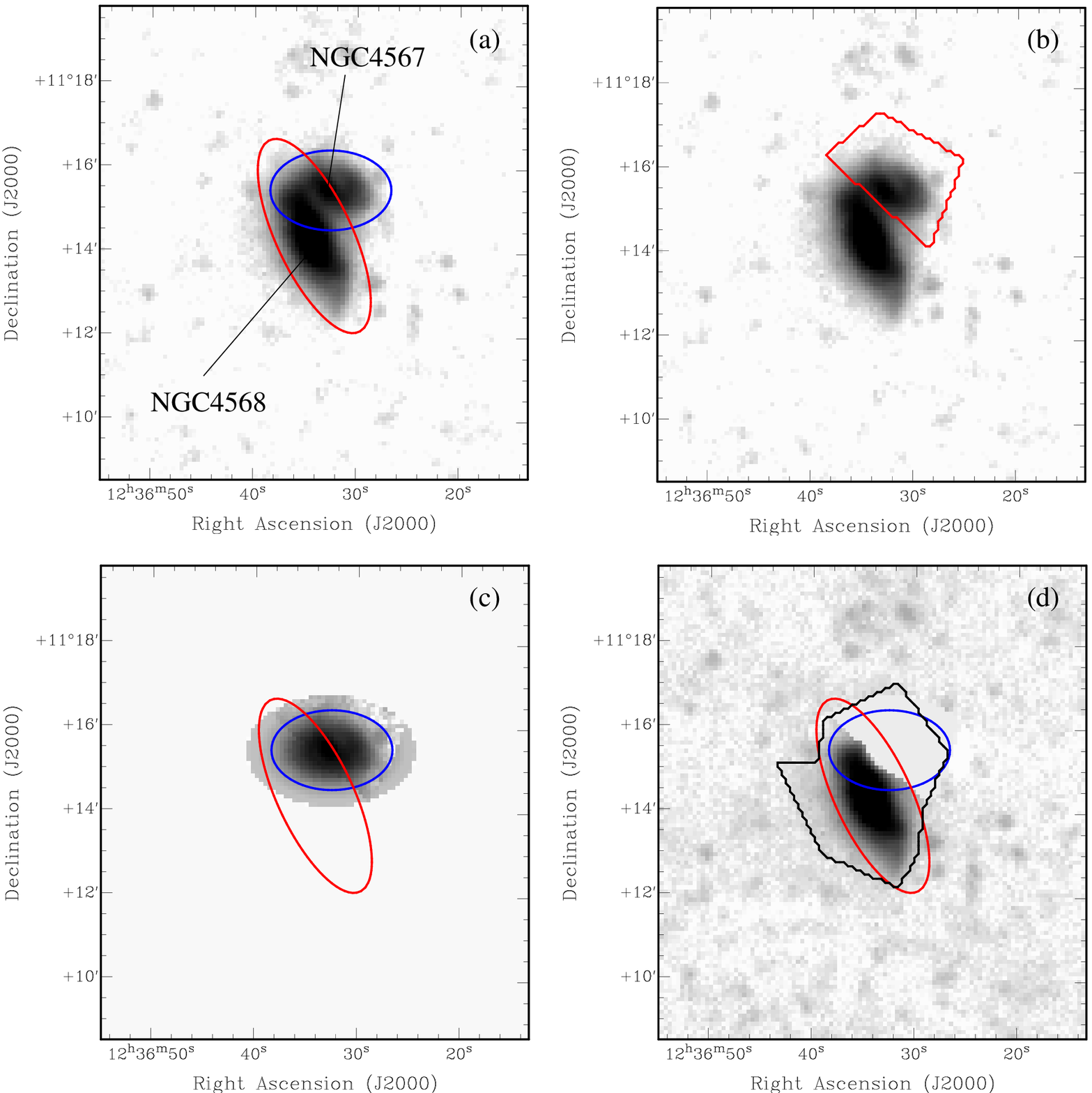}
	\caption{Images of the modelling process for galaxy pair NGC4567/8. Log flux scales have been employed to enhance the bright and dim features. (a) 250 \micron\ greyscale image showing the two galaxies and their optical extent measured to $d_{25}$. (b) Isolated emission from NGC4567 (red outline) was used to create a mask. (c) the model of NGC4567. (d) The model-subtracted image showing only the emission from NGC4568. The aperture that was used to measure the total flux of the system is shown outlined in black.\label{fig7}}
\end{center}
\end{figure*}

The interaction between NGC4567 and NGC4568 gives rise to a region of FIR emission consisting of contributions from each galaxy. Fig. \ref{fig7}a shows the 250 micron map of the region with the $d_{25}$ ellipses outlined. A mask was constructed from the 250 \micron\ data to isolate the emission from only NGC4567 (Fig. \ref{fig7}b). This galaxy was chosen because it is the fainter of the two. Since it is fainter, any error in the NGC4567 model will have less impact on the flux estimate in the overlapping region than attempting to model NGC4568. 

The ellipse fitting procedure described in Section \ref{dr} was applied to the masked 250 \micron\ image. The procedure was manually limited to an extent of 2\arcmin - the minimum distance to cover the overlapping region. The model was then constructed in two parts; for the region containing emission from only NGC4567, the original map data were used, for the overlap region, the surface brightness values from the ellipse fitting were used (Fig. \ref{fig7}c). 

The model was then subtracted from the original map, to produce a map with only the estimated flux from NGC4568. An aperture was defined to encompass both galaxies in the original map to enable us to measure the total flux for both galaxies combined. Care was also taken to exclude bright background galaxies from this aperture. The same aperture was then used to measure the estimated flux of NGC4568 in the model-subtracted map (Fig. \ref{fig7}d). 

The estimate of the flux in NGC4567 was then simply taken as the difference of the two measurements. The procedure was then repeated using the 250 \micron\ mask and aperture as the templates for the other bands. The resulting fluxes are very similar to the BGS values which employed a simple spatial cut between galaxies to define the extent of each. This is probably because the emission in the overlap region is dominated by NGC4568. We have adopted a conservative estimate for the flux measurement uncertainty of 20\% across all bands.

\clearpage

\section{Table of FIR fluxes and upper limits for 750 VCC galaxies in the HeViCS field}
\label{fluxtbl}
\centering
\begin{longtable}{c@{ }c@{ }c@{ }c@{}c@{}c@{}c@{}c@{}c@{}} 
\hline
OBJECT & RA & dec. & Type & $S_{500}$ ($\sigma S_{500}$) & $S_{350}$ ($\sigma S_{350}$) & $S_{250}$ ($\sigma S_{250}$) & $S_{160}$ ($\sigma S_{160}$) & S$_{100}$ ($\sigma S_{100}$)\\
&h:m:s&d:m:s&&(Jy)&(Jy)&(Jy)&(Jy)&(Jy)\\
&(J2000)&(J2000)&&&&&&\\
\hline
  VCC1 & 12:08:20.30 & +13:41:02.0 & BCD & $<$ (0.021) & 0.029 (0.005) & 0.088 (0.008) & 0.179 (0.033) & 0.198 (0.031)\\
  VCC4 & 12:08:31.60 & +15:05:42.0 & Im (Im/S) & $<$ (0.025) & $<$ (0.024) & $<$ (0.011) & $<$ (0.011) & $<$ (0.011)\\
  VCC9 & 12:09:22.30 & +13:59:33.0 & dE (d:E) & $<$ (0.027) & $<$ (0.017) & $<$ (0.011) & $<$ (0.015) & $<$ (0.010)\\
  VCC10 & 12:09:24.90 & +13:34:28.0 & BCD & 0.065 (0.019) & 0.161 (0.021) & 0.326 (0.032) & 0.417 (0.065) & 0.398 (0.058)\\
  VCC11 & 12:09:35.80 & +06:44:36.0 & dE (d:E) & $<$ (0.012)* & $<$ (0.019) & $<$ (0.012) & ---  & --- \\
\end{longtable}

\clearpage

\section{Table of dust masses and temperature derived from single-temperature modified blackbody fits to 140 VCC galaxies}

\label{temptbl}
\centering
\begin{longtable}{cccccccc} 
\hline
Object & RA & dec & Type & Distance (Mpc) & $T$ (K) & log($M_{D}$/M$_{\odot}$) & $\chi^2_{dof=3}$\\
\hline
  VCC10 & 12:09:24.90 & +13:34:28.0 & BCD & 32 & 19.8 (1.0) & 6.4 (0.1) & 3.3637\\
  VCC25 & 12:10:37.50 & +16:01:59.0 & Sc & 32 & 22.5 (0.8) & 7.5 (0.1) & 1.3392\\
  VCC28 & 12:10:45.60 & +15:51:55.0 & Sc & 17 & 17.2 (1.1) & 5.7 (0.2) & 2.6363\\
  VCC47 & 12:12:11.80 & +13:14:48.0 & Sa & 32 & 17.6 (0.6) & 7.0 (0.1) & 3.4034\\
  VCC58 & 12:12:32.30 & +12:07:26.0 & Sb & 32 & 17.5 (0.6) & 7.4 (0.1) & 2.584\\
  VCC73 & 12:13:03.00 & +07:02:20.0 & Sb & 32 & 23.7 (0.8) & 7.3 (0.1) & 0.9731\\
  VCC89 & 12:13:47.30 & +13:25:29.0 & Sc & 32 & 21.5 (0.7) & 7.6 (0.1) & 1.6889\\
\hline
\end{longtable}

\clearpage
\section{FIR SED plots with single-temperature modified blackbody fits}
\label{onetemps}

\begin{figure*}
\centering
	\includegraphics[height=0.83\textheight]{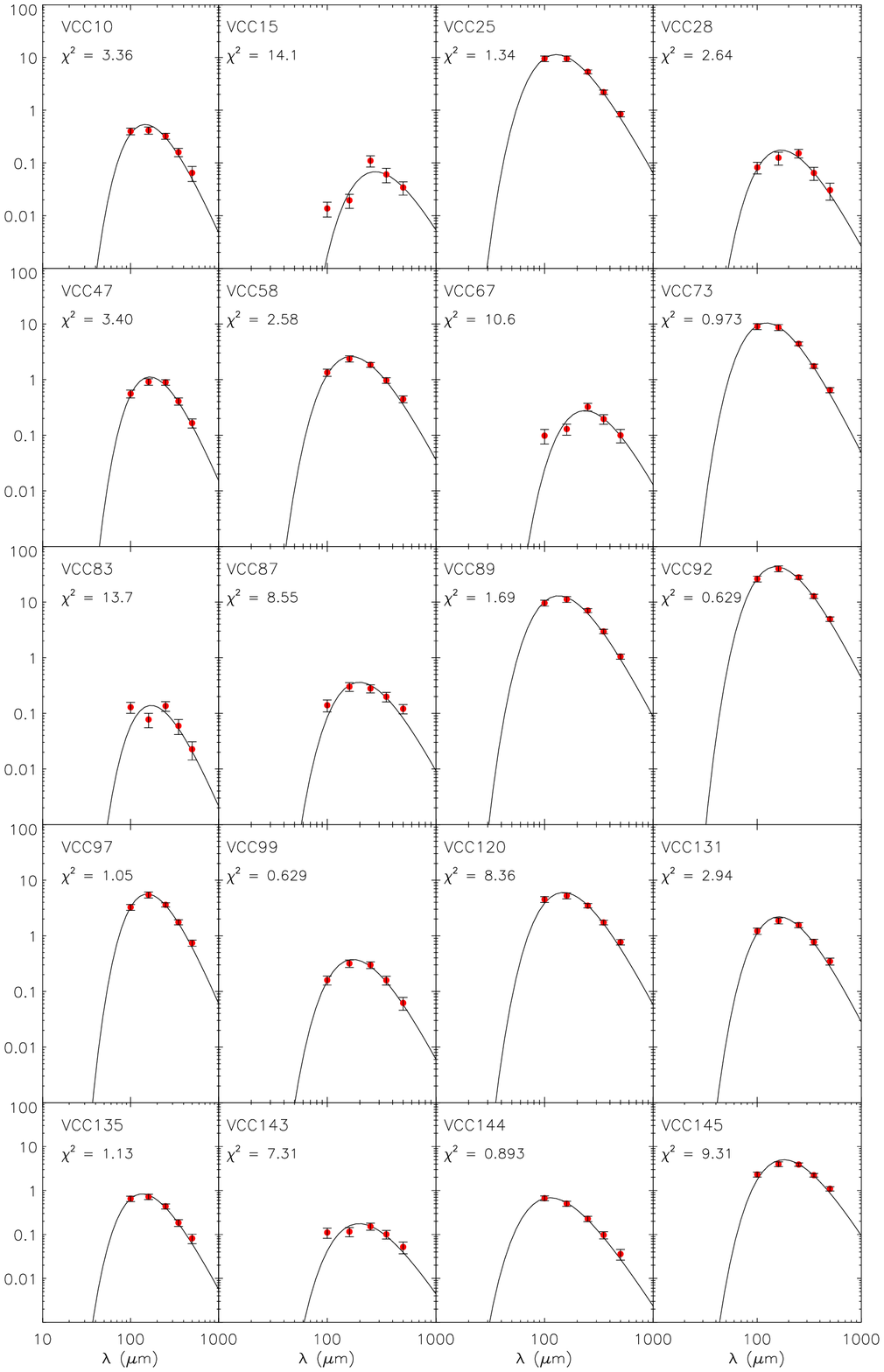}
	\caption{FIR SED plots for the VCC galaxies with 5-band FIR detections. Also shown are the single-temperature modified blackbody fits (solid black line).}
	\label{fig6a}
\end{figure*}

\begin{figure*}
\centering
	\includegraphics[height=0.9\textheight]{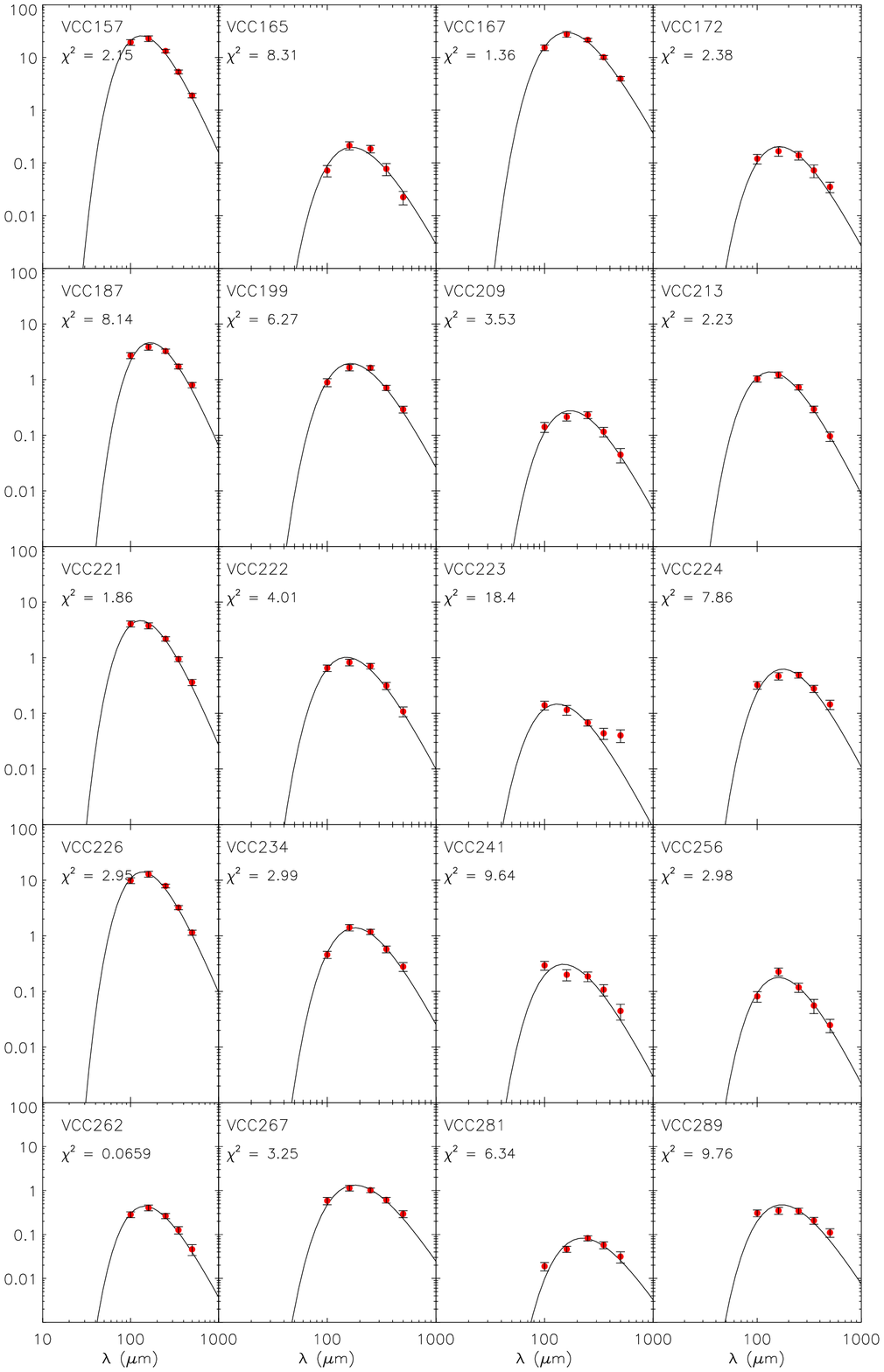}
	\contcaption{FIR SED plots with single-temperature modified blackbody fits}
	\label{fig6b}
\end{figure*}

\begin{figure*}
\centering
	\includegraphics[height=0.9\textheight]{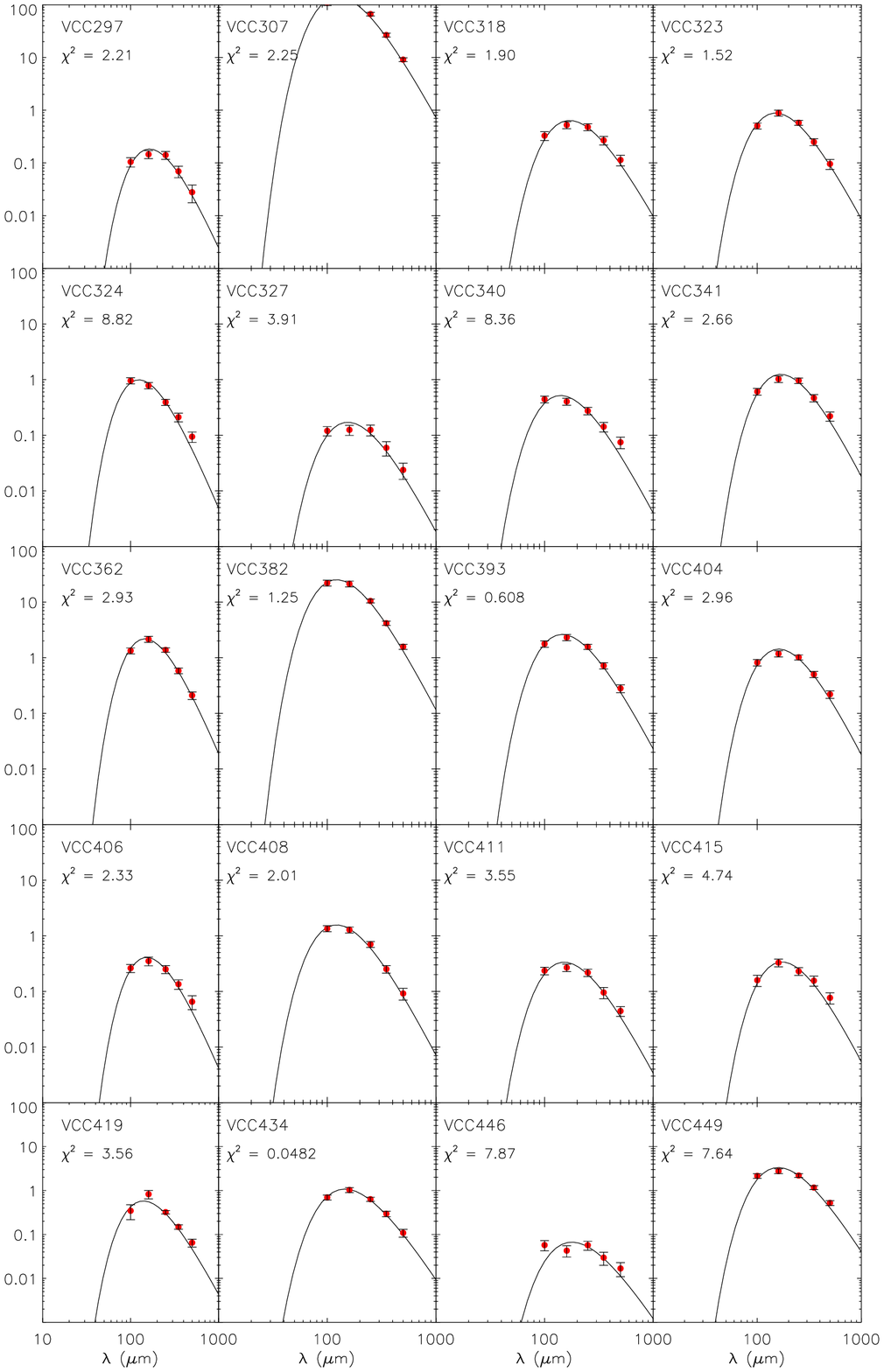}
	\contcaption{FIR SED plots with single-temperature modified blackbody fits}
	\label{fig6c}
\end{figure*}

\begin{figure*}
\centering
	\includegraphics[height=0.9\textheight]{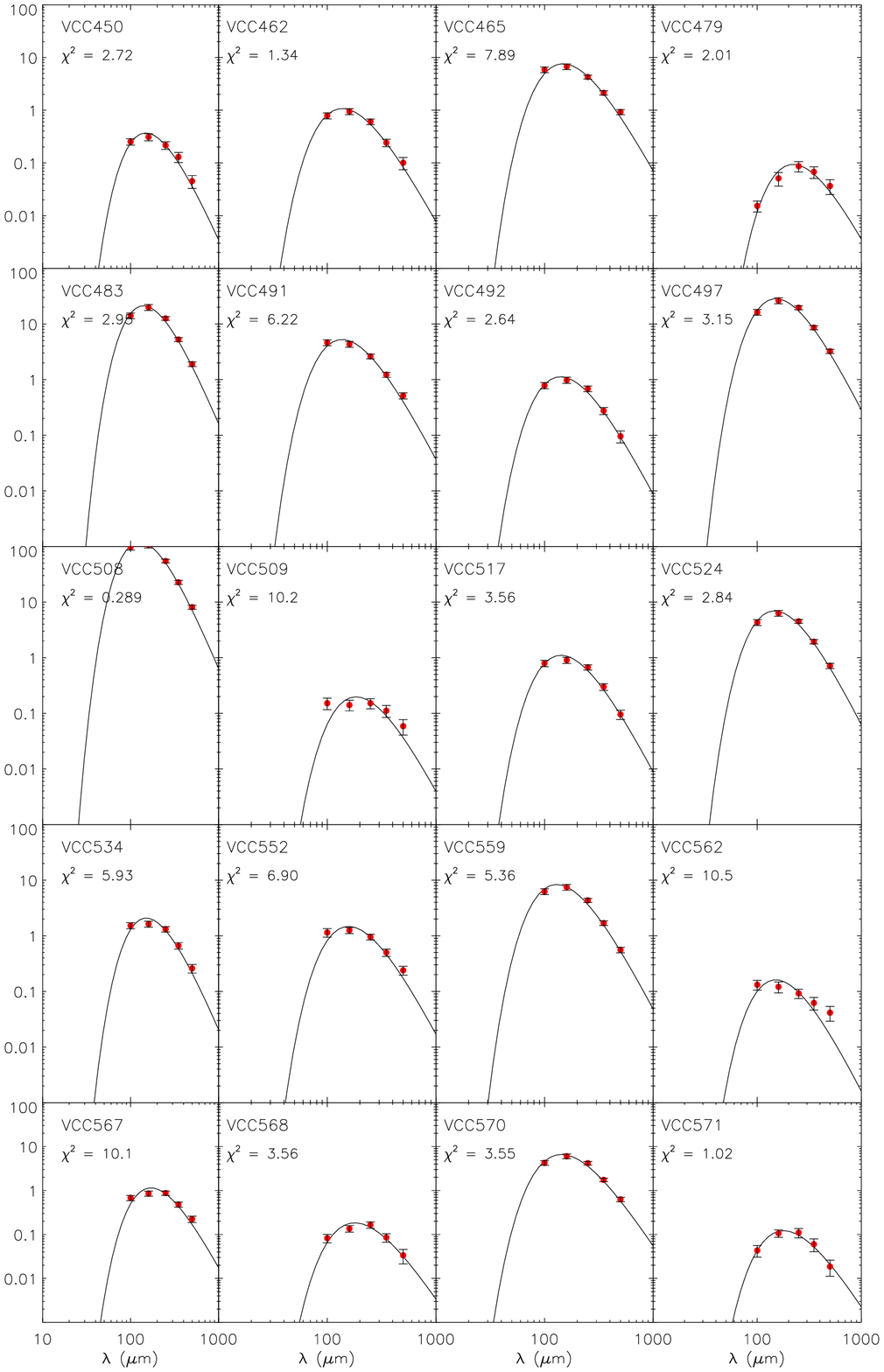}
	\contcaption{FIR SED plots with single-temperature modified blackbody fits}
	\label{fig6d}
\end{figure*}

\begin{figure*}
\centering
	\includegraphics[height=0.9\textheight]{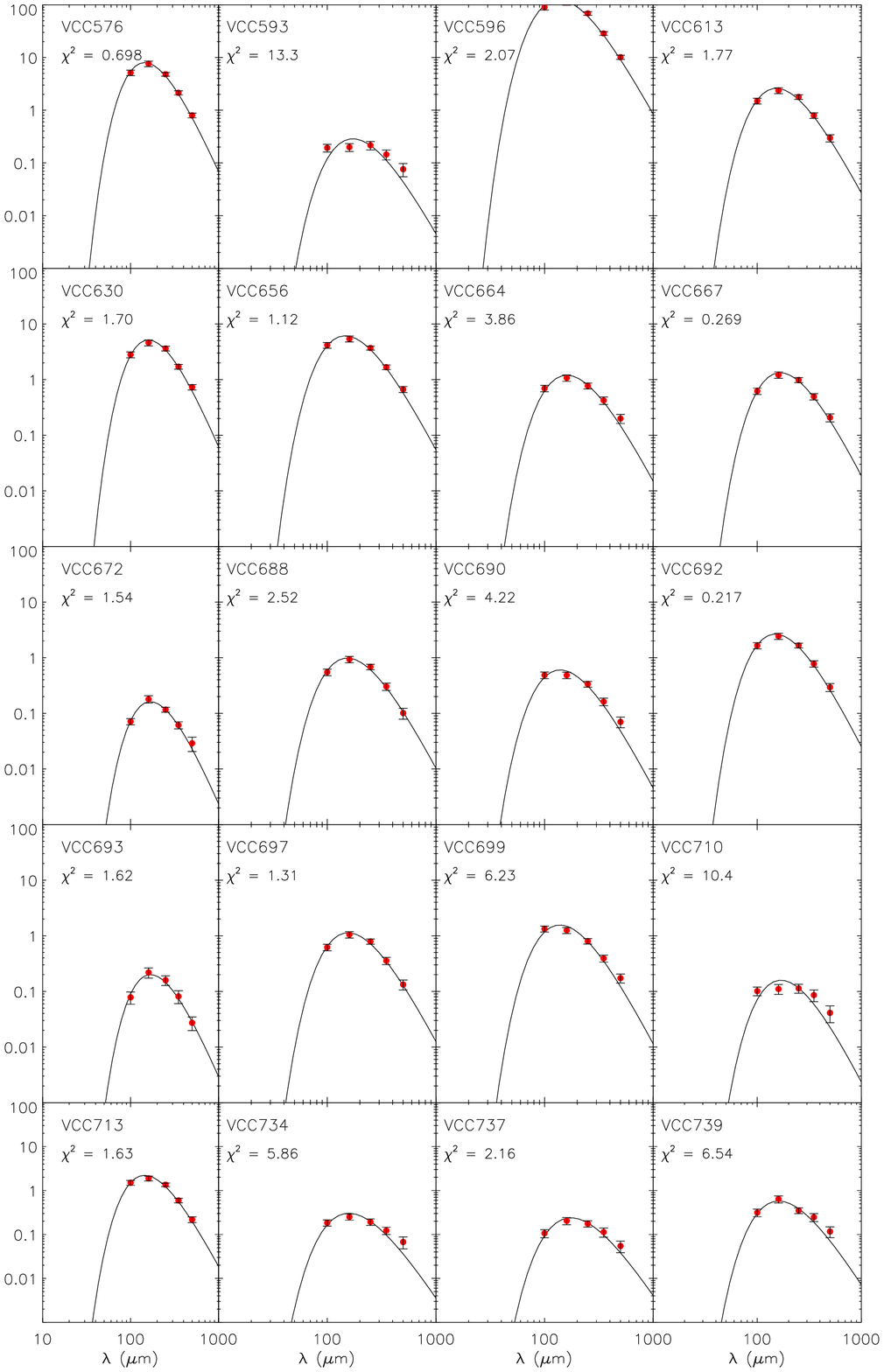}
	\contcaption{FIR SED plots with single-temperature modified blackbody fits}
	\label{fig6e}
\end{figure*}

\begin{figure*}
\centering
	\includegraphics[height=0.9\textheight]{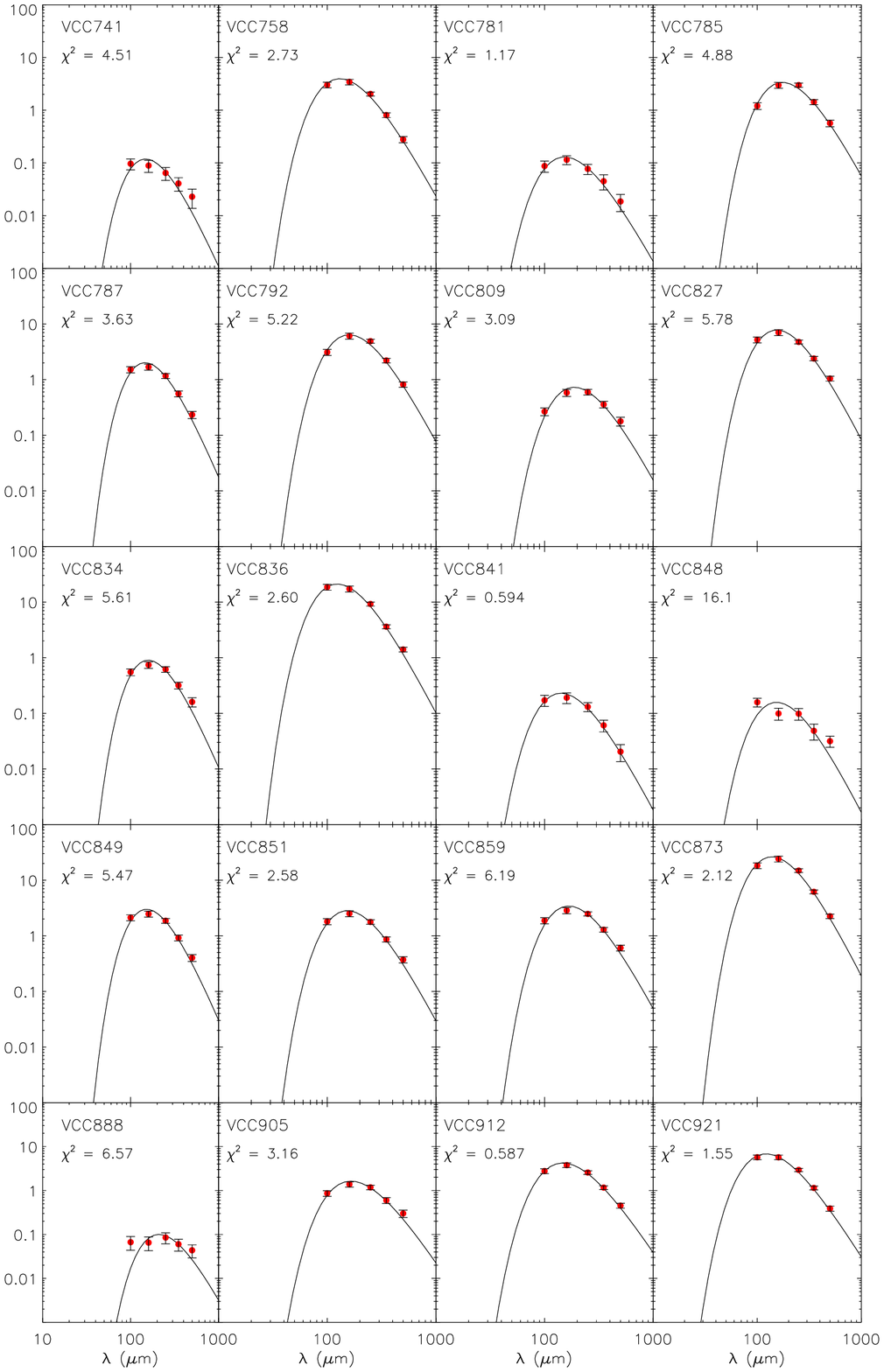}
	\contcaption{FIR SED plots with single-temperature modified blackbody fits}
	\label{fig6f}
\end{figure*}

\begin{figure*}
\centering
	\includegraphics[height=0.9\textheight]{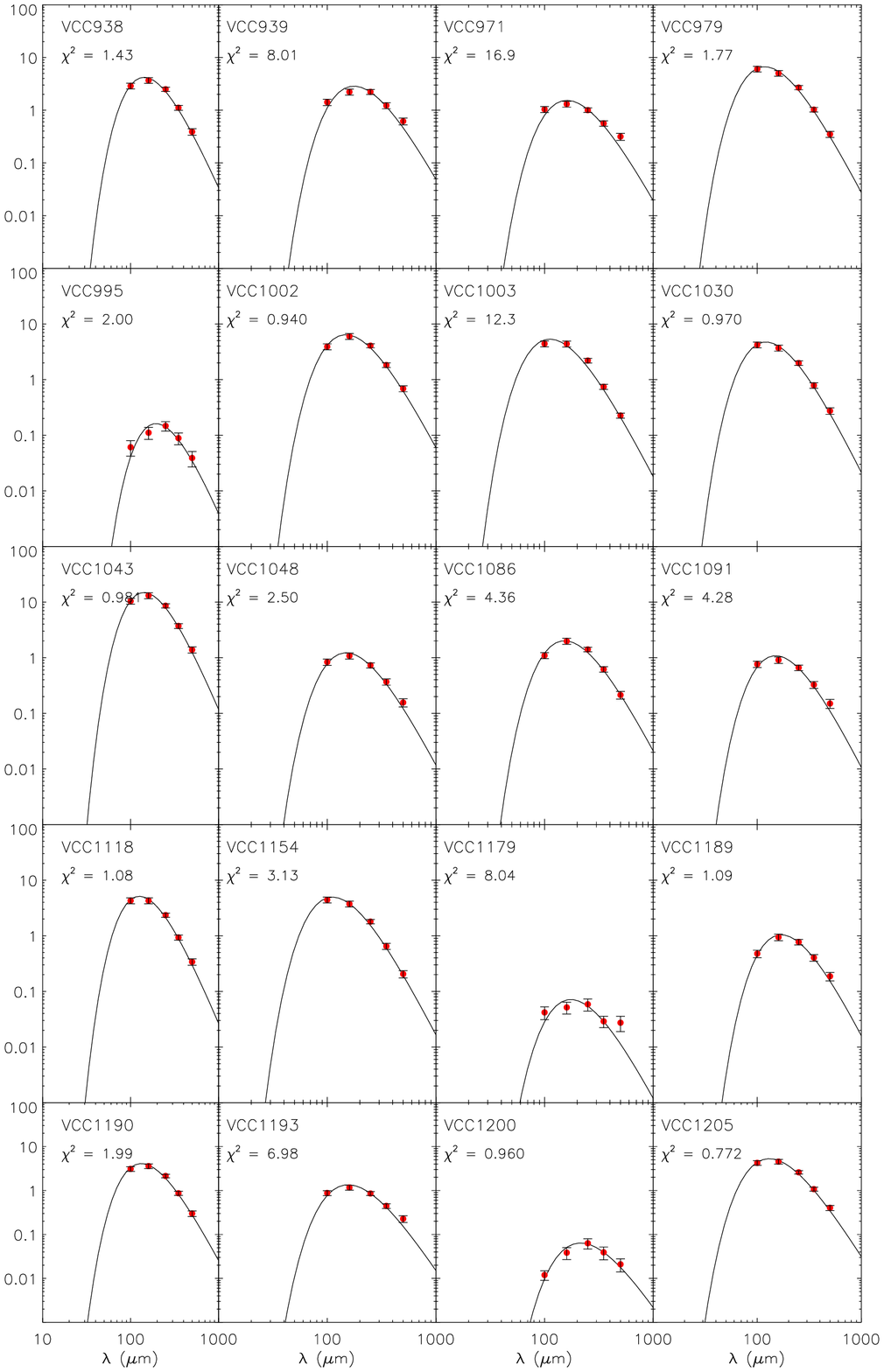}
	\contcaption{FIR SED plots with single-temperature modified blackbody fits}
	\label{fig6g}
\end{figure*}

\begin{figure*}
\centering
	\includegraphics[height=0.9\textheight]{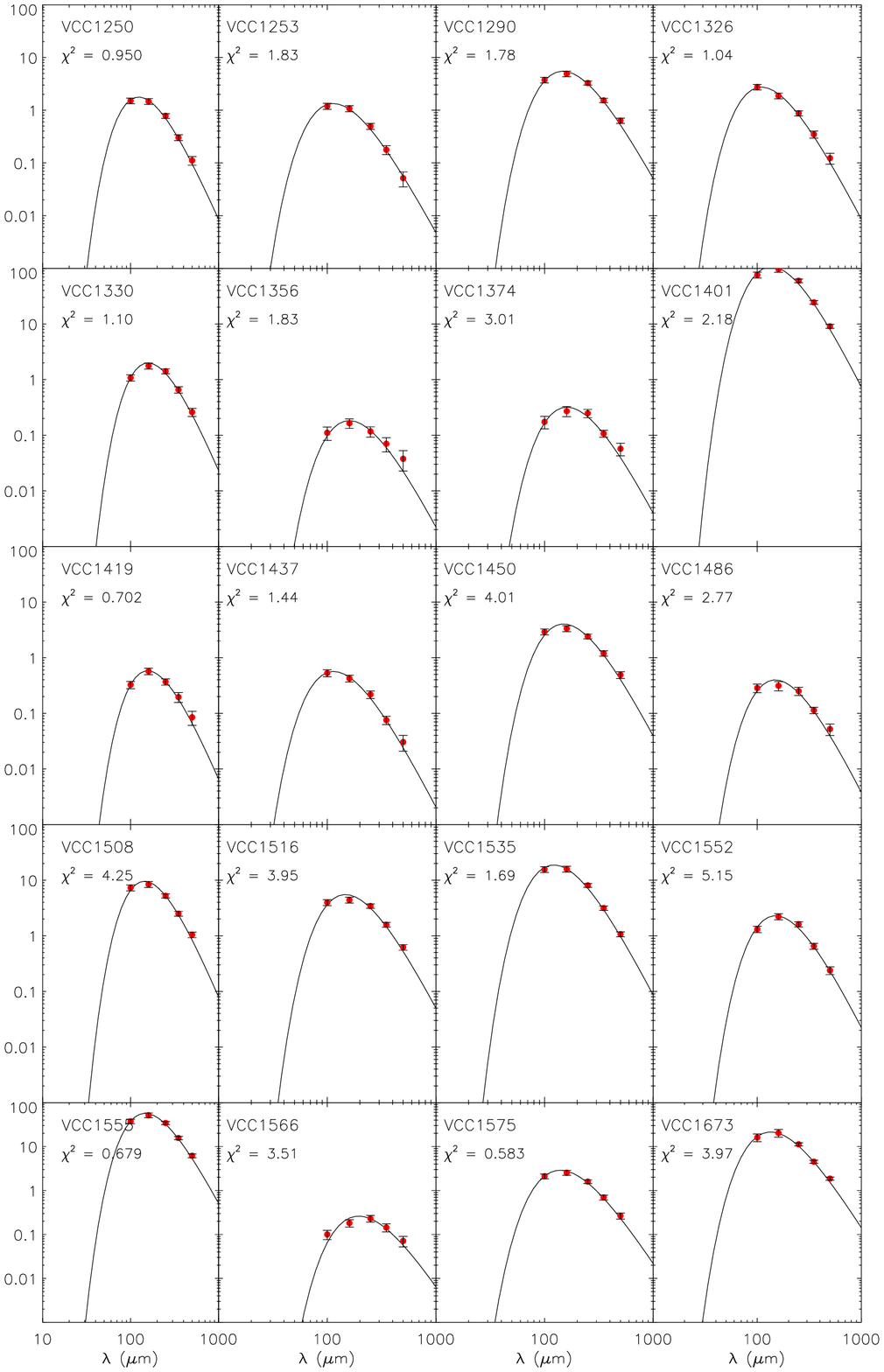}
	\contcaption{FIR SED plots with single-temperature modified blackbody fits}
	\label{fig6h}
\end{figure*}

\begin{figure*}
\centering
	\includegraphics[height=0.9\textheight]{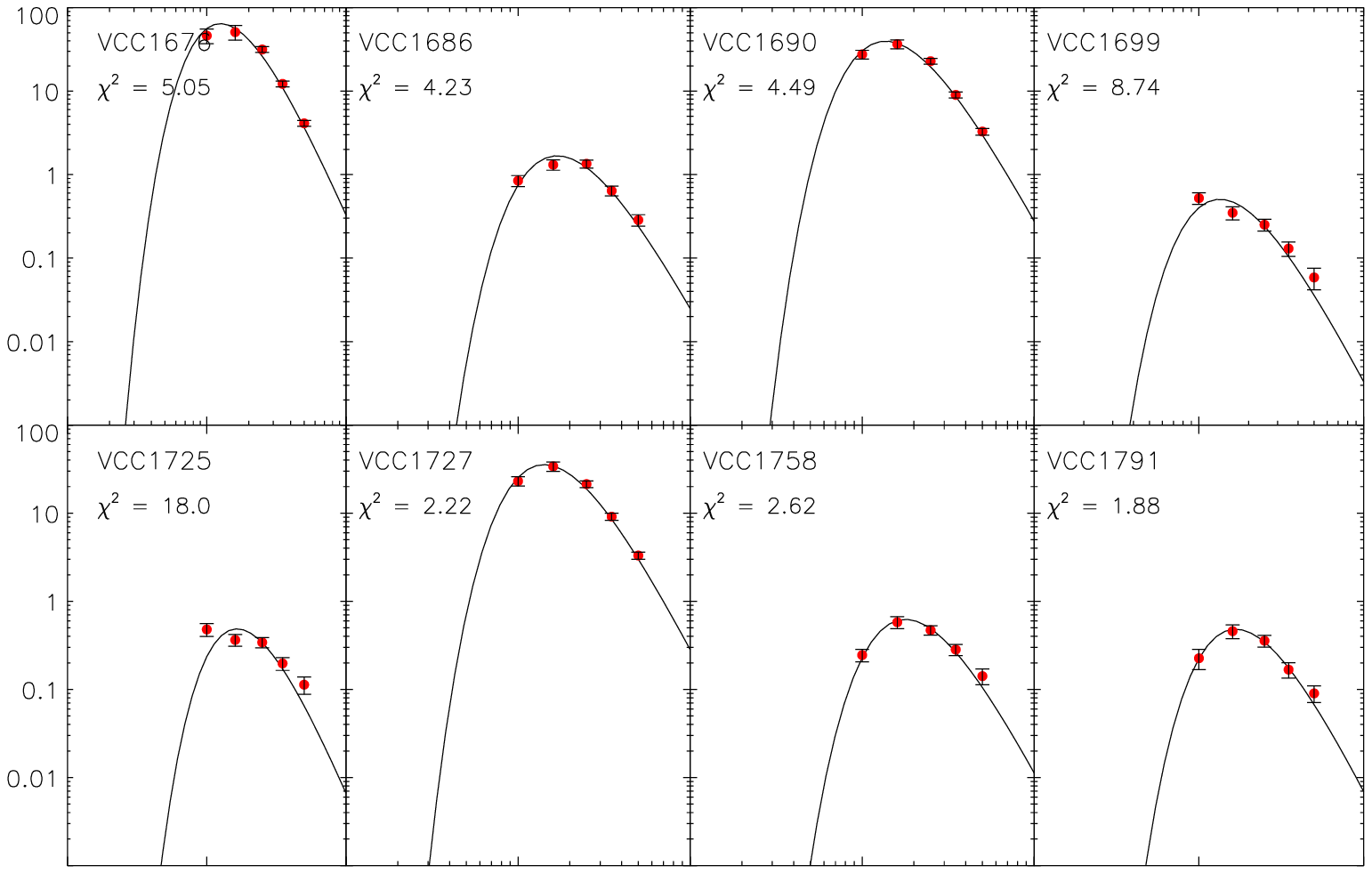}
	\contcaption{FIR SED plots with single-temperature modified blackbody fits}
	\label{fig6i}
\end{figure*}

\clearpage

\twocolumn

\end{document}